\documentclass{ar-1col}
\usepackage{enumerate}
\usepackage{amssymb}
\usepackage{natbib}
\usepackage{color}
\usepackage{comment}
\jname{Xxx. Xxx. Xxx. Xxx.}
\jvol{AA}
\jyear{2018}
\doi{10.1146/((please add article doi))}
\newcommand{\gtsim}{\lower.5ex\hbox{$\; \buildrel > \over \sim \;$}} 
\newcommand{\ltsim}{\lower.5ex\hbox{$\; \buildrel < \over \sim \;$}} 
\newcommand\arcdeg{\mbox{$^\circ$}} 
 
\newcommand\arcsec{\mbox{$^{\prime\prime}$}} 
\newcommand\degr{\arcdeg} 
\begin{document}
\markboth{Roger Blandford, David Meier, Anthony Readhead}{AGN Jets}
\title{Relativistic Jets in Active Galactic Nuclei}
\author{Roger Blandford$^1$, David Meier$^2$, and Anthony Readhead$^3$.
\affil{$^1$KIPAC, Stanford University, Stanford, CA 94305, USA: rdb3@stanford.edu}
\affil{$^2$Dept. Astronomy, Caltech, Pasadena, CA 91125 \& Jet Propulsion Laboratory, Pasadena, 91109}
\affil{$^3$Owens Valley Radio Observatory, Caltech, Pasadena, CA 91125}}
\begin{abstract}
The nuclei of most normal galaxies contain supermassive black holes, which can accrete gas through a disk and become active. These Active Galactic Nuclei, AGN, can form jets which are observed on scales from AU to Mpc and from meter wavelengths to TeV gamma energies. High resolution radio imaging and multi-wavelength/messenger campaigns are elucidating the conditions under which this happens.  

Evidence is presented that:
\begin{itemize}
\item AGN jets are formed when the black hole spins and the accretion\\disk is strongly magnetized, perhaps on account of gas accreting\\at high latitude beyond the black hole sphere of influence.
\item AGN jets are collimated close to the black hole by magnetic stress\\associated with a disk wind.
\item Higher power jets can emerge from their galactic nuclei in a\\relativistic, supersonic and proton-dominated state and they\\terminate in strong, hot spot shocks; lower power jets are degraded\\to  buoyant plumes and bubbles. 
\item Jets may accelerate protons to EeV energies which contribute to\\the cosmic ray spectrum and which may initiate pair cascades\\that can efficiently radiate synchrotron gamma rays.
\item Jets were far more common when the universe was a few billion\\years old and black holes and massive galaxies were growing rapidly. 
\item Jets can have a major influence on their environments, stimulating\\and limiting the growth of galaxies. 
\end{itemize}
The observational prospects for securing our understanding of AGN jets are bright.
\end{abstract}
\maketitle
\tableofcontents

\section{INTRODUCTION}\label{sec:intro}
\begin{extract}
``A curious straight ray lies in a gap in the nebulosity in p.a. $20\degr$, apparently connected with the nucleus by a thin line of matter. The ray is brightest at its inner end, which is 11$\arcsec$ from the nucleus.'' -- Heber D. Curtis description of a $5\,{\rm min}$ exposure of NGC4486 (M87), 1918.
\end{extract}
\subsection{Historical Context}\label{ssec:hist}
As the above quotation attests, jets were first observed over a century ago.  Much has been learned since then through the application of new technology, which enabled remarkable gains in sensitivity and angular and spectral resolution and which has revealed jets throughout the entire electromagnetic spectrum as well as the gravitational, and perhaps neutrino and cosmic ray, windows. Simultaneously, understanding of relativity, quantum mechanics and the physics of ionized gas provided a secure foundation to interpret these observations.

\citet{fath09} produced the first manifestation of an AGN
\begin{marginnote}[-2pt]\entry{AGN}{Active Galactic Nucleus. Compact region at the center of a galaxy with high, non-stellar luminosity}\end{marginnote}
but it was not until \citet{seyfert43} that the presence of a central gravitational well was suggested, in retrospect, by the observation of broad emission lines from gas moving at high speed. Radio astronomy began in the 1930s \citep{jansky33,reber40}. A key discovery was the resolution of a bright radio source, Cygnus A, into two components or lobes \citep{jennison53} straddling a galaxy at a redshift $z=0.05$ \citep{baade54}.  Synchrotron radiation was identified as the radio emission mechanism and shown to require enormous energies \citep{shklovsky55,burbidge56}. The first quasar, 3C273, was discovered in 1963 \citep{schmidt63,hazard63} and shown to have a radio power ($L_{\rm rad}\equiv dL/d\ln\nu(6\,{\rm GHz})$), with $L$ the luminosity, even greater than Cygnus A.
\begin{marginnote}[2pt]\entry{Quasar}{An AGN that outshines its host galaxy}\end{marginnote}
It had a compact flat radio spectrum core, with spectral index $\alpha\equiv d\ln L/d\ln\nu\sim0$ and a linear jet extending to 20 arcseconds, very similar to that in M87. The spectrum of 3C273 was distinguished from that of Cygnus A by the presence of broad, as opposed to narrow, emission lines, which is a distinguishing feature between radio galaxies and quasars. 
\begin{marginnote}[-2pt]\entry{RLQ}{Radio-Loud Quasar. Quasar with $10^{40}\,{\rm erg\,s}^{-1}<L_{\rm rad}\lesssim10^{46}\,{\rm erg\,s}^{-1}$}\end{marginnote}
Another distinguishing feature is that the radio sources associated with RLQs are often compact, whereas those associated with radio galaxies are extended, although most RLQ large enough to be resolved show the same double-lobe morphology as radio galaxies. 

AGN were observed throughout the electromagnetic spectrum. The historical optical magnitude and the contemporary radio flux of 3C273 varied dramatically on timescales of days to years \citep{smith63,dent65}.  More generally, blazars were found to be highly variable in the radio \citep{hughes65}, optical \citep{schmidt63}, X-ray \citep{schreier82} and $\gamma$-ray \citep{bignami81,punch92} bands, on timescales that can be as short as minutes. 
\begin{marginnote}[2pt]\entry{Blazar}{AGN with relativistic jet directed towards us}\end{marginnote} RQQ were found in optical surveys and shown to be roughly ten times numerous than RLQ and just as powerful in the optical \citep{sandage65}. 
\begin{marginnote}[-2pt]\entry{RQQ}{Radio-Quiet Quasar. Quasar with $10^{38}\,{\rm erg\,s}^{-1}\lesssim L_{\rm rad}<10^{40}\,{\rm erg\,s}^{-1}$}\end{marginnote}
Before long it became clear that most normal galaxies have nuclei which can be identified through their spectral lines, stellar activity or nonthermal emission \citep{keel83}. This includes our own Galaxy, for which the nucleus Sgr A$^\ast$, \citep{balick74} has a radio luminosity $L_{\rm rad}\sim10^{33}\,{\rm erg\,s}^{-1}\sim10^{-3}L_{\rm bol}$, with $L_{\rm bol}$ being the bolometric luminosity.

Theory, viewed selectively, kept up with these remarkable discoveries. Some early models of quasars \citep{zeldovich64,salpeter64} correctly invoked SMBH largely to account for the high radiative efficiency. 
\begin{marginnote}[2pt]\entry{SMBH}{Super Massive Black Hole. Spinning black hole with mass in the range $\sim10^{6-10}\,{\rm M}_\odot$}\end{marginnote}
Accretion disks \citep{lyndenbell69,bardeen70} were investigated, and double radio sources were seen to be powered continuously by jets emanating from galactic nuclei \citep{rees71}, rather than as remnants of single explosions like supernovae.  Their rapid radio variability led \citet{rees66} to propose relativistic motion in RLQ. This suggestion -- that relativistic motion, which with its concomitant angular beaming effects dominates the observed characteristics of RLQ -- was the first critical step towards unifying FSRQ with radio galaxies (mostly steep spectra). 
\begin{marginnote}[2pt]\entry{FSRQ}{Flat Spectrum Radio Quasar. High power blazar with quasar properties}\end{marginnote}
It also alleviated the physical challenges posed by compact radio sources \citep{hoyle66}. Generic particle, fluid and electromagnetic models of jets were explored to account for their origin, collimation and radiation \citep{scheuer74,blandford74}.

\subsection{This Review}\label{ssec:rev}
In preparing this review on AGN jets, we have chosen to emphasize topics where there has been recent progress or where we judge it to be imminent. Much of this activity involves multi-wavelength and messenger investigations of AGN in general. The research literature is therefore immense --- maybe a thousand citable references per year--- and we have had to make ruthless choices to keep this review within editorial bounds. We have chosen to emphasize papers that either give a clear context to the research, provide instructive examples of more general jet properties,  or connect well to upcoming investigations. We eschew references where extensive discussion can be easily found on websites. Consequently, many important technical advances, discoveries, ideas and detailed studies can only be recognized implicitly. To those involved, we offer sincere apologies.

The following section contains an overview of direct observations of jets together with relevant discussion of the black holes and accretion disks that launch them and the radio lobes that they feed. \textbf{Section~\ref{sec:kindyn}} deals with kinematic and dynamical inferences that have been drawn from these observations. Jets are mostly observed nonthermally, which requires relativistic particles to be accelerated and radiation mechanisms to be identified. We discuss these matters in Section~\ref{sec:emmod} and turn to the relationship of jets to AGN and the universe at large in \textbf{Section~\ref{sec:uni}}. Finally, in \textbf{Section~\ref{sec:sum}}, we summarize our view of what is reasonably well-established about AGN jets and list some important open issues that should be understood over the next decade using new telescopes.

\section{OBSERVATIONS OF RELATIVISTIC JETS}\label{sec:obs}
\begin{extract}
``We shall call this `super-light expansion.'' -- from \citet{cohen71}.
\end{extract}
\subsection{Background}\label{ssec:radio}
We can now count $\sim10^9$ extragalactic radio sources, mostly with $L_{\rm rad}\gtrsim10^{38}\,{\rm erg\,s}^{-1}$ (comparable with the Milky Way), in low frequency radio surveys. A substantial fraction of these sources are cores, jets and lobes powered by SMBH; the remainder are dominated by stellar processes. \citep{padovani16}.  The observation and theory of SMBH sources were reviewed by \citet{bridle84,begelman84,rees84}. While there has been much progress and many fine reviews since, the early results provide a convenient starting point for much of our discussion.

\subsubsection{Unified Theories}\label{sssec:unif}The early radio surveys were all carried out at frequencies of a few hundred MHz, and although individual sources were followed up at higher frequencies the low frequency surveys led to a strong focus on steep spectrum sources, since these objects predominate at low frequencies.  In a seminal paper \citet{kellermann69} drew attention to the importance of flat spectrum sources, but it was not until the S4 survey \citep{ptoth78} produced a complete sample of 269 radio sources at 5 GHz that the importance of the flat-spectrum sources was fully recognized. $\sim 60$\% of the strong ($>0.6$ Jy) radio sources selected at 5 GHz are compact ($<1$ arc second) flat spectrum objects, predominantly quasars, while $\sim40$\% are extended ($>1$ arc second) steep spectrum objects, predominantly galaxies \citep{ptoth78} . This was an important result because it provided a key ingredient to theories that unified radio galaxies and quasars. In one of the largest radio galaxies, NGC6251, the pc-scale jet was found to be parallel to, and pointing in the same direction as, the 200 kpc jet \citep{readhead78a}. This result, coupled with the flat/steep spectrum  radio source dichotomy, and superluminal motion, led \citet{readhead78b} to suggest relativistic beaming (see \textbf{Figure~\ref{fig:m87mon}}) and orientation \citep{rees66} as the possible reasons for the apparent differences at radio wavelengths between these two classes --- {\it viz.} flat/steep spectra, compact/extended structure, misaligned/aligned with larger scale structure, superluminal/absence of superluminal motion (\textbf{Sidebar Special Relativistic Effects}).
\begin{marginnote}[2pt]\entry{BLL}{BL Lac object. Low power blazar that lacks prominent broad emission lines}\end{marginnote}
\citet{blandford78} extended these arguments to BLL, and \citet{blandford79,konigl81} presented theoretical discussions of relativistic jets as compact radio, X-ray and $\gamma$-ray sources. Over the next two years the results continued to strengthen the case for unified theories  \citep{readhead80, orr82}. The results favoring ``a powerful unifying theory'' were summarized in \citet{begelman84},  but there was one major piece of the puzzle that did not fit --- the strong broad optical emission lines in quasars {\it vs.} the weak narrow emission lines in radio galaxies. This last major piece of the puzzle was solved when NGC1068 was shown to have broad polarized emission lines, providing the first direct evidence of equatorial rings of dusty gas surrounding AGN, which can obscure the broad line region \citep{antonucci85}.
 
\begin{textbox}[ht]
\section{SPECIAL RELATIVISTIC EFFECTS}
\subsection{Superluminal Expansion}
Suppose that radiation is emitted by a moving source S at time $t_{\rm S}$ from position ${\bf r}(t_{\rm S})$ along direction $\bf n$ to an observer at large distance $d$ who receives it at a time $t_{\rm O}=t_{\rm S}-{\bf n}\cdot{\bf r}/c$ after the time that a pulse would have been received from the origin of the coordinate system. Ignoring cosmological expansion, the apparent speed of S observed by O is $v_{\rm apparent}=V\sin\theta/(1-V\cos\theta/c)$, where $V=|\dot{\bf r}|$ and $\cos\theta={\bf n}\cdot\dot{\bf r}/V$. This can be superluminal and has a maximum value $\Gamma V$, with the Lorentz factor $\Gamma=(1-V^2/c^2)^{-1/2}$ when $V=c\cos\theta$. Typically $\Gamma\sim10$. If S is a physical source (and this may not be the case), the Doppler factor associated with S is given by ${\cal D}\equiv\nu_{\rm O}/\nu_{\rm S}=T_{\rm B\,O}/T_{\rm B\,S}=[\Gamma(1-V\cos\theta)]^{-1}$, where $\nu$ is the frequency and $T_{\rm B} \equiv Sc^2/2k\nu^2\Omega$ is the brightness temperature for a source with flux density $S$ subtending an angular size $\Omega$.
\subsection{Doppler Boosting}
If the radiation has an intensity (flux per Hz  per sterad) $I$, then this transforms according to $I(\nu_{\rm O})={\cal D}^3I(\nu_{\rm S})$. The effect can be dramatic. The flux observed from an approaching, flat spectrum ($\theta=\alpha=0$), optically thin source with $\Gamma\sim10$ is larger than that from an identical receding source by $\sim6\times10^7$. The intensity is determined by solving the equation of radiative transfer in the galaxy frame transforming the emissivity from a frame moving with S to the O frame using $j(\nu_{\rm O})={\cal D}^2I(\nu_{\rm S})$. Likewise, the absorption coefficient transforms according to $\mu(\nu_{\rm O})={\cal D}^{-1}\mu(\nu_{\rm S})$.
\end{textbox}

\begin{table}[ht]
\tabcolsep7.5pt
\caption{Classification of AGN by their Optical Properties} 
\begin{center}
\begin{tabular}{@{}l|c|c|c@{}}
\hline
Abbrev.&Expansion&Definition&Density ($L_{\rm bol}$)$^{\rm a}$\\
\hline\hline
LINER &Low-Ionization Nuclear& Weak Seyfert-like galaxy& $\sim10^{6.5}$ ($<10^{42}$)\\&Emission-line Region & & \\
\hline
Sy 2&Seyfert galaxy type 2&AGN with narrow permitted&$\sim10^{5.3}$ ($>10^{42}$)\\& & \& forbidden lines & \\
\hline
Sy 1 & Seyfert galaxy type 1 & AGN with broad permitted & $\sim10^{5}$ ($>10^{42}$)\\& & \& narrow forbidden lines & \\
\hline
QSO & Quasi-Stellar Object & Powerful AGN that &$\sim10^{2.5}$ ($>10^{45}$)\\& & outshines its host galaxy & \\
\hline\hline
WLRG& Weak-Line Radio Galaxy & radio galaxy analog to LINER & $\sim10^{4}$ ($<10^{42}$)\\
\hline
NLRG& Narrow-Line Radio Galaxy & radio galaxy analog to Sy 2 & $\sim10^{1.2}$ ($>10^{42}$)\\
\hline
BLRG& Broad-Line Radio Galaxy & radio galaxy analog to Sy 1 & $\sim10$ ($>10^{42}$)\\
\hline
QSR & Quasi-Stellar Radio source & QSO with strong radio emission &$\sim10^{-1.5}$ ($>10^{45}$)\\
\hline
\end{tabular}
\end{center}
\begin{tabnote}
$^{\rm a}$ Gpc$^{-3}$, local; AGN $L_{\rm bol}$ in erg $s^{-1}$, \citet{tadhunter16}; bright field galaxy density $\sim10^{7}$ 
\end{tabnote}
\label{tab:ROQ}
\end{table}

\subsection{Radio Jets}
\subsubsection{Three Important Jet Scales}\label{sssec:scales}AGN jets exist on scales from solar system to galaxy separation scales, from $\lesssim1\,{\rm AU}$ to $\gtrsim1\,{\rm Mpc}$ and are now observed on $\mu as$ to degree scales using single dishes, Long and Very Long Baseline Arrays. Reviews of the first three decades of VLBI surveys, and VLBI studies of pc-scale jets, are given by \citet{wilkinson95}, and \citet{zensus97}, respectively. 
\begin{marginnote}[2pt]\entry{VLBI}{Very Long Baseline Interferometry}\end{marginnote}
More recently \citet{beasley02} led the VLBA Calibrator Survey, which has been augmented by  a number of additional surveys including observations of southern sources made with the Australian Long Baseline Array \citep{petrov11}.  Thus far, more than 12,000 AGN have been observed by VLBI and these form the basis of the Astrogeo Project \citep{petrov17}. Other recent extensive VLBI studies of pc-scale jets are those of 295 AGN by the MOJAVE group \citep{lister16a}, and the monthly 43 GHz VLBI observations of 34 $\gamma-$ray bright blazars  \citep{marscher11, jorstad16}.

AGN jets can be considered on three, loosely-defined scales. Most observations are of Galaxy Jets where the dynamical environment is dictated by the stellar/dark matter potential within the host galaxy and the interstellar medium. Black Hole Jets are the inward extension of galaxy jets to the gravitational radius of the black hole, $r_{\rm g} \equiv {GM/c^2}$ (\textbf{Sidebar Black Holes}). Here, the environment is dominated by the black hole potential and the inflow onto an accretion disk and the outflow from it (\textbf{Sidebar Black Hole Accretion}). These two regimes straddle the radius of influence of the SMBH $R_{\rm inf}=GM/\sigma^2$, where $\sigma$ is the 1D central stellar velocity dispersion and of order the \citet{bondi52} radius, which is a measure of the distance at which the gravitational potential changes abruptly and from which gas can accrete onto the hole. Lobe jets are the outward extension of galaxy jets, from $\sim0.1R_{\rm lobe}$ to $R_{\rm lobe}$ where $R_{\rm lobe}$ is the maximal extent of the jet or lobe. Here, the environment is controlled by the backflow from the end of the jet and the circumgalactic medium.

\subsubsection{Galaxy Jets ($R_{\rm inf}\lesssim R\lesssim0.1R_{\rm lobe})$}\label{sssec:galjets}
\paragraph{Structure and Kinematics}\label{par:stkin}
Three key observational facts emerged in the early years of VLBI imaging: (i) they are one-sided jets \citep{wrpa77};(ii) they have a compact flat-spectrum core at one end of a steep-spectrum jet \citep{readhead78b}; and (iii) components are often seen to expand or to move along the jet away from the core at superluminal speeds \citep{gubbay69, zensus87}.
  
By far the most comprehensive VLBI study monitoring blazars is that of the MOJAVE group \citep{lister16a, lister16b}, where the results of VLBI monitoring observations of 400 AGN jets spanning 20 years are presented. Almost all objects observed have the same one-sided core-jet structure, as was seen in the earliest VLBI maps, and relativistic motion is usually observed with $v_{\rm apparent}$ ranging from $\sim 0.03c$ to 40 c. (In one object (PKS 0805-07), $v_{\rm apparent}$ reached $\sim 50$c in 1996 before decelerating to $\sim 20$c.) Superluminal motion is common in FSRQs, BLL, and Narrow Line Seyfert 1 galaxies.  Acceleration is observed in many superluminal components, and 32\% of the jet features display non-radial motion while 4\% showed inward motion. Almost all the AGN with the fastest moving components have been detected by Fermi Gamma-ray Space Telescope, henceforth Fermi, indicating a strong correlation between bulk Lorentz factor and $\gamma-$ray emission.  Within the first $\sim 100$ pc the majority of the features observed are superluminal and are accelerating.

\begin{textbox}[ht]
\section{BLACK HOLES}
Einstein's general theory of relativity has been abundantly corroborated in the weak field limit and its strong field, nonlinear version is consistent with many observations involving cosmology, gravitational radiation and black holes \citep{meier12,thorne17}. Black holes are now seen as inevitable consequences of stellar and galactic evolution. Massive black holes with masses $M\sim10^{6-10}{\rm M}_\odot$ appear to be present in the nuclei of most normal galaxies, including our own. Examination of a remarkable solution \citep{kerr63} for the metric of a spinning black hole shows that there is an event horizon with radius (measured by the circumference) $r_+=r_{\rm g}[1+(1-j^2)^{1/2}]$. $r_{\rm g} \equiv GM/c^2=1.5(M/10^6\ {\rm M}_\odot) \times 10^{6} {\rm km}$ is the ``gravitational radius'' and $\tau_g \equiv r_{\rm g} /c=5(M/10^6\ {\rm M}_\odot)\ {\rm s}$ is the gravitational time. $j\equiv Jc/(GM^2)$ is the dimensionless spin angular momentum of the hole with $|j|<1$. The angular velocity of the hole is $\Omega_{\rm H}=jc/2r_+$. Many spin measurements are close to maximal \citep{reynolds14}. 
\end{textbox}

\begin{figure}[hb]
\includegraphics[width=5in]{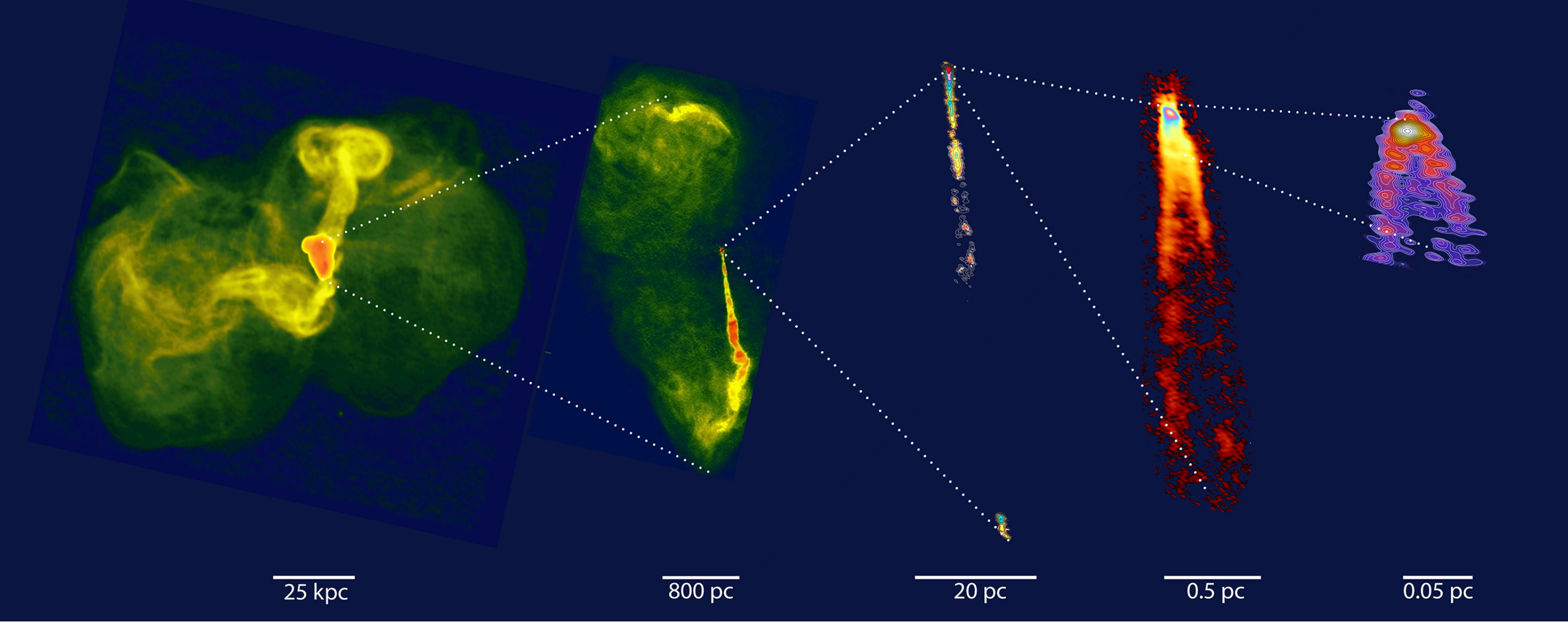}
\caption{Montage of the FR-I radio galaxy M87, on scales from the outer lobes to near the black hole. From left:  (a) Lobe jet and outer lobes, showing present outburst and a more ancient one almost perpendicular to the former; (b) galaxy jet and inner lobes; (c) full view of the black hole jet, including HST-1 at $R_{\rm inf}$ (bottom); (d) innermost jet; (e) jet launching region near SMBH. The last four images show the striking effects of relativistic beaming, even with the  jet pointing at a modest $17\degr$ to our line-of-sight \citep{walker18}. The counter-jet is largely invisible, pointing away from us at $166\degr$. Credit: (a) NRAO, 90 cm VLA; (b) NRAO, 20 cm VLA; (c) NRAO: 20 cm VLBA \citep{cheung07}; (d) NRAO, 7 mm VLBA \citep{walker18}, (e) 3 mm global VLBI network \citep{kim18a}} 
\label{fig:m87mon}
\end{figure}

Recent advances in mm VLBI have been reviewed by \citet{boccardi17},  together  with the roles of the global VLBI network and the present as well as the anticipated role of the EHT.
\begin{marginnote}[2pt]\entry{EHT}{Event Horizon Telescope}\end{marginnote}
The 43 GHz VLBI observations of M87 \citep{mertens16} show clearly a mildly relativistic velocity component along the sheath and a faster component $\Gamma=2.5$ along the spine. The global mm VLBI network has also been used at 86 GHz to study the jet of M87 down to $\sim7\,r_{\rm g}$ \citep{kim18a} and shows limb brightening, and hence evidence of a spine-sheath structure down to this scale, which therefore appears to be anchored to the inner portion of the accretion disk.  The result  of stacked images from this study is shown in \textbf{Figure~\ref{fig:m87mon}}.

\begin{figure}[ht]
\includegraphics[width=4in]{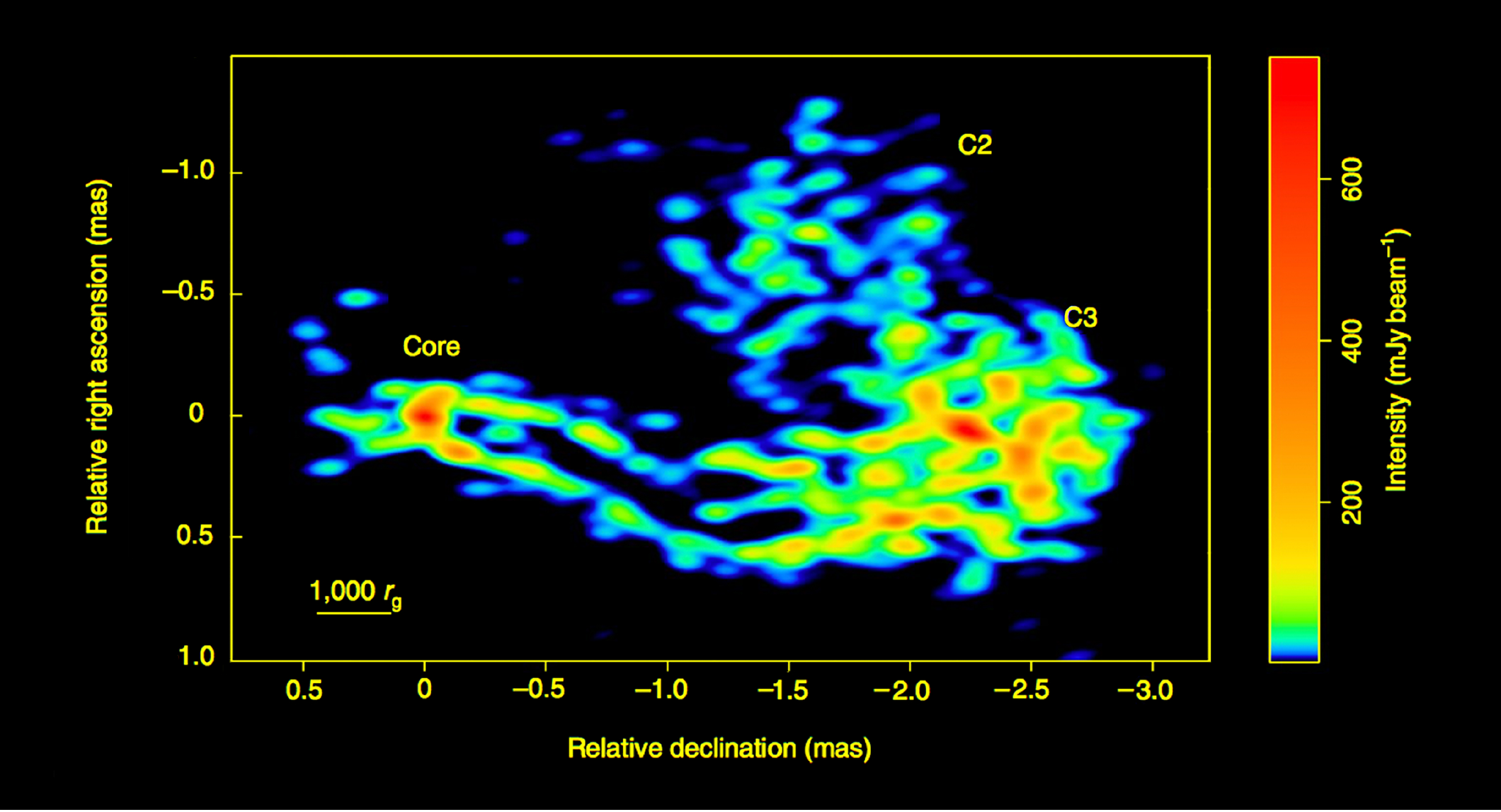}
\caption{22 GHz image of the black hole jet in 3C84 made with the global VLBI network and RadioAstron space VLBI antenna \citep{giovannini18}.  With a viewing angle of $\sim 18 \degr$, the deprojected distance from core down to component C3 is $\sim10^{4.5} \, r_{\rm{g}}$ or $\sim3$ pc. Rather than collimating in a semi-parabolic fashion, like M87, the black hole jet in 3C84 appears to have a nearly uniform width from near the core region to a distance of $\sim 10^4 r_{\rm g}$.}
\label{fig:3c84}
\end{figure}

The  very  high  resolution  22  GHz  observations  of  3C84  made  with  the  global  VLBI network and RadioAstron space VLBI mission (\citet{giovannini18}, \textbf{Figure~\ref{fig:3c84}}) show an edge-brightened jet down to within 30 $\mu$as from the core, and they conclude that the jet either goes through a very rapid lateral expansion on scales $\lesssim100 r_{\rm g}$ or is launched from the accretion disk.  The global mm VLBI network has been used at 86 GHz to study the polarization of 3C84 by \citet{kim18b}, who find that the polarization is consistent with an underlying limb-brightened structure, and they find, due to its uniform RM structure, that the Faraday depolarization is most likely due to an external screen.

\begin{textbox}[ht]
\section{BLACK HOLE ACCRETION}
\subsection{Accretion Disks}
AGN black holes are surrounded by orbiting gas. When this gas is able to cool on the inflow timescale, it will form a thin accretion disk in the equatorial plane with an inner radius radius $r>r_{\rm ISCO}$, the radius of the Innermost Stable Circular Orbit which shrinks from $9r_{\rm g}$ to $r_{\rm g}$ as $j$ increases from $-1$ to $1$ \citep{meier12}. The outer radius of the disk is unknown; self-gravitation, dust and even stars can be important \citep{thompson05}. In this simple model, gas spirals inward under magnetic torques sustained by the Magneto-Rotational Instability (MRI) \citep{balbus98} and plunges towards the event horizon within $r_{\rm ISCO}$. 

However, the gas will not cool when the dimensionless accretion rate $\dot{m} \gtrsim 1$, where $\dot{m} \equiv \dot{M} / (4\pi GMm_p/\sigma_Tc)$ is scaled to the Eddington rate, as the radiation is trapped by the inflowing gas and radiation pressure thickens the disk into a torus which can create a funnel which may be responsible for the initial collimation of radio quasar jets \citep{abramowicz13}. A torus supported by the pressure of hot ions can perform a similar function when the accretion rate is so low that electrons can remain mildly relativistic on a flow timescale \citep{yuan14}. 

\subsection{Classical and Revisionist Accretion Disk Theory}
Initially, interiors of black hole accretion disks were modelled like those of stars. At very low $\dot{m}$ the disk was thought to be a geometrically thin, optically thick hot ionized plasma supported by gas pressure and the opacity given by free-free absorption, similar to low-mass stars \citep{shakura73}. Such disk inflow is called ``outer region inflow''. For higher rates, in the range $10^{-3} \sim \dot{m}_M \lesssim \dot{m} \lesssim \dot{m}_I \sim 0.2 \, m^{-1/8}$, ``middle region inflow'' occurs, with the opacity near the black hole dominated by Thompson electron scattering, and the  gas pressure still dominant. For even higher rates, in the range $\dot{m}_I \lesssim \dot{m} \lesssim 1$, radiation pressure dominates the ``inner region inflow''. In the early scaling models, inner region flow was found to be secularly and thermally unstable, but numerical simulations of the process are still inconclusive. For $\dot{m} \gtrsim 1$ radiation pressure should drive a strong outflowing wind.

However, at low accretion rates ($\dot{m} \lesssim \dot{m}_A \sim 0.05 - 0.1$) black hole accretion may be more like a geometrically {\em thick}, optically {\em thin} stellar corona described by a very hot, two-temperature ``advection-dominated accretion flow'' (ADAF) with ions at $\sim 10^{12} \rm{K}$ supplying the pressure and electrons at $\sim 10^9-10^{11} \rm{K}$ producing most of the emitting radiation. Note that, at accretion rates below $\dot{m}_A$, outer region inflow simply does not exist as $\dot{m}_A > \dot{m}_M$ is always the case. Similarly, even middle region inflow should not exist for massive black holes ($m > 10^{3 - 5}$), as $\dot{m}_A > \dot{m}_I$ because of the mass dependence of $\dot{m}_I$. So AGN might be expected to have only {\em three} types of accretion flow:  ADAF at low rates, unstable inner region flow at high rates, and outflow at super-Eddington rates. An alternative view \citep{blandford99} of low rate accretion is that the torque always does so much work on the outer disk that it is unavoidably unbound and once the inflow is unable to radiate the binding energy released, there will be an outflow so that the mass flow into the black hole is much less than the mass supply rate at large radius. If the outflow is strongly magnetized, the disk may remain thin at all radii \citep{konigl94}.
\end{textbox}

\paragraph{Polarization}\label{par:inpol}
A very intensive linear polarization study was carried out  on 484 AGN using observations from the MOJAVE program as well as from the VLBA archive, covering 20 years from 1996-2016 \citep{pushkarev17}. All told 5410 VLBA observations were used in this study. The principal clear results to emerge were the following: (i) for all classes fractional polarization increased with core separation; (ii) a clear increase of polarization fraction towards the edge of the jet, interpreted as being due to the fact that the greater depth of the jet along lines of site closer to the axis of the jet, compared to the edge of the jet,  leads to more Faraday depolarization; (iii) 40\% of jet cores have a preferred EVPA
\begin{marginnote}[2pt]\entry{EVPA}{Electric Vector Position Angle}\end{marginnote}
direction from epoch to epoch; (iv) BLL cores have more stable EVPAs than quasars and tend to be aligned with the initial jet direction; (v) BLL jets show the same tendency; (vi) quasars and radio galaxies show no such tendency. 

A review of the full Stokes imaging results, \textit{i.e.} including circular polarization,   from the MOJAVE program is given in \citet{homan18},
presenting multi-epoch results from 2002 to 2009, for 278 objects over 6 epochs on average. Typical levels of circular polarization, when detected, range from 0.3\% to 0.7\%, with the maximum observed being 1\% in NRAO 140, which was also fairly stable, unlike most sources.  The maximum circular polarization is always seen in the core component, and the large majority showed a preferred sign, which persisted for 3.5 years -- i.e. longer than the duration of a typical flare.  3C279 has been observed for  more than 20 epochs spanning 14 years all with the same sign of circular polarization.

 Faraday rotation measure studies, reviewed by \citet{gabuzda17}, have been pursued intensively by a number of groups. \citet{zavala04} made a systematic study of the RM
\begin{marginnote}[2pt]\entry{RM}{Rotation Measure}\end{marginnote}
in 40 quasars, radio galaxies, and BLL objects, and found that the RMs in  both quasars and BLL are very similar in both the cores  (typically 500 to several thousand rad ${\rm m}^{-2}$), and  the jets (typically 500  rad ${\rm m}^{-2}$ or less). In contrast, the cores of radio galaxies are generally unpolarized, whereas the jets exhibit RMs ranging from a few hundred to $\sim 10^4$  rad ${\rm m}^{-2}$. Their results can best be explained in terms of a screen in close proximity to the jet, which could very well indicate the presence of  a sheath around the jet. Magnetic fields  $\sim 1 \; \mu$G are deduced in the screen.

The MOJAVE group has carried out a systematic study of Faraday rotation in 191 extragalactic radio jets, using multifrequency VLBA observations over 12 epochs, and found that quasars have on  average larger RMs than BLL. Of particular interest is the fact that they observed transverse gradients in RM \citep{hovatta12} in four blazars including 3C273, in which the RM changes sign over the transverse cuts, which they interpret as evidence for a helical magnetic field. This result has been confirmed  \citep{wardle18}, and provides firm evidence of a torroidal magnetic field component, which requires a current down the jet of $10^{17}$--$10^{18}$ A.

A very interesting discovery in optical polarization observations of AGN is that large rapid rotations of the EVPA of some AGN are correlated with $\gamma-$ray flares \citep{marscher08}.  While an intrinsic connection between rotations and gamma-ray flares is difficult to establish on an event-by-event basis, due to uncertainties on both the gamma-ray and polarimetry side, a more general connection between optical polarization and gamma-ray activity has now been firmly established by the RoboPol program \citep{blinov18}. 

\subsubsection{Black Hole Jets ($r_{\rm g}\lesssim R\lesssim R_{\rm inf}$)}\label{sssec:inner}
\paragraph{M87 Imaging}\label{par:m87}
Most images we have of black hole jets are associated with lower power, FR-I, objects -- M87, BL Lac and 3C84. Because of the proximity of M87 (16 Mpc), VLBI imaging of its jet affords a very detailed look at what lies inside the ``core'' region ($R < 10^6 \, r_{\rm g}$). This includes the ACZ.
\begin{marginnote}[2pt]\entry{ACZ}{Acceleration and Collimation Zone}\end{marginnote}
The most detailed and systematic VLBI study of the innermost M87 jet carried out so far is that of \citet{walker18}, which is based on 50 VLBA 43 GHz observations with a resolution of $\sim 60\times120\;r_{\rm g}$, taken over 17 years. These observations show the following (\textbf{Figure~\ref{fig:m87mon}}): (i) an asymmetric jet and counter-jet in the inner 1.5 mas (0.12 pc); (ii) both jet and counter-jet are edge brightened; (iii) both jet and counter-jet show an initial rapid widening followed by a narrowing and then a second widening at which point the counter-jet becomes invisible; (iv) the jet is subsequently collimated; (v) proper motions and the intensity ratios of the jet and counter-jet indicate acceleration from $v_{\rm apparent}<0.5 c$ to $v_{\rm apparent}>2 c$; and (vi) polarization observations suggest a helical magnetic field close to the core. 

\begin{figure}[ht]
\includegraphics[width=4in]{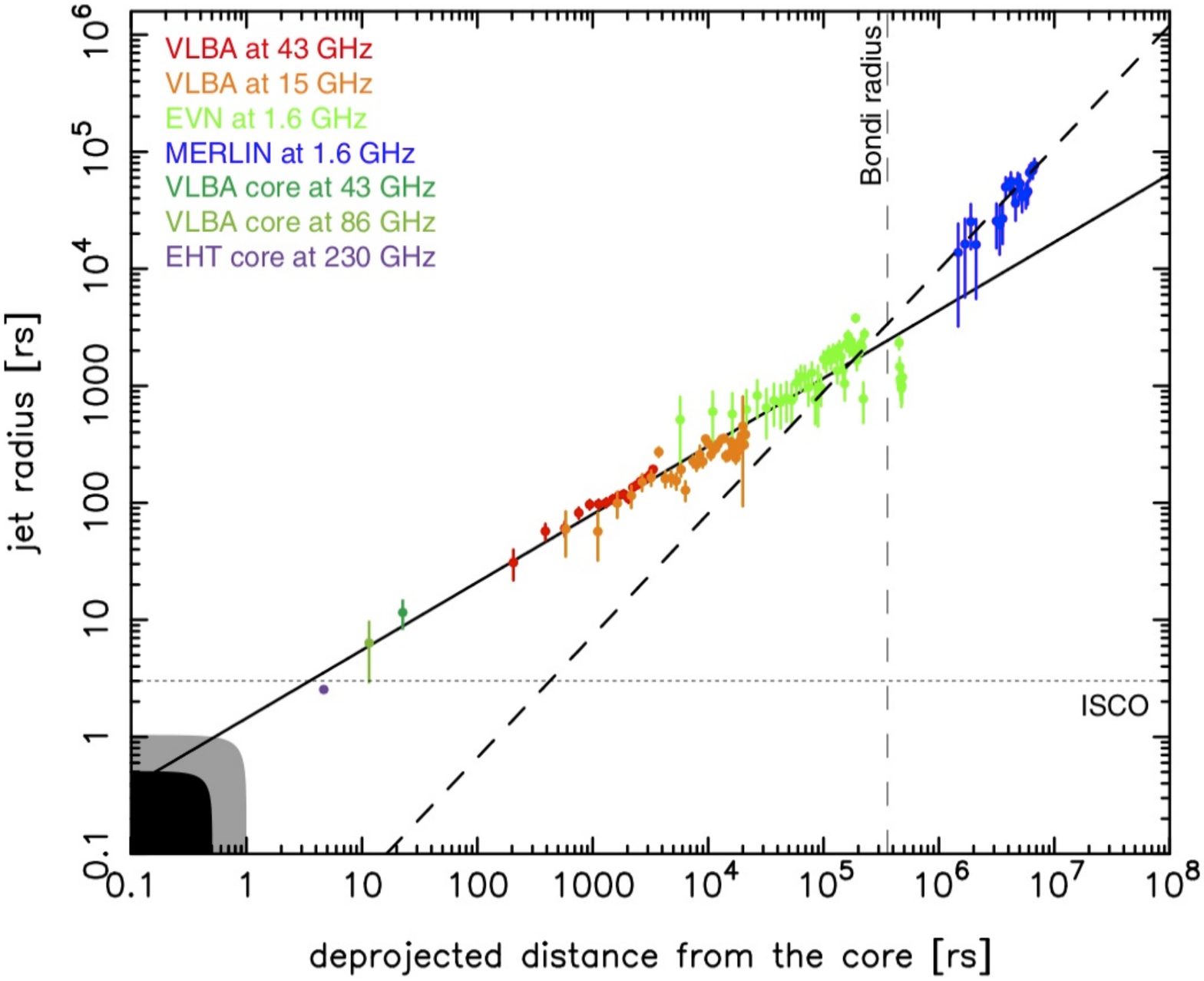}
\caption{Transverse radius \textit{vs.} $R$ for the black hole and galaxy jets in M87 (see third and fourth images in \textbf{Figure~\ref{fig:m87mon}}).  The jet collimates (and accelerates) in a semi-parabolic fashion steadily from the black hole to a maximum apparent velocity of $\sim6 c$ at HST-1 ($\sim4 \times 10^5{\rm r_{\rm S}} \approx 10^6 {\rm r_{\rm g}} \,\sim R_{\rm inf}$), where there is a change in slope or jet collimation break. Beyond that the jet flares in a linear fashion and decelerates to sub-relativistic speeds \citep{asada12}. The Bondi radius is about $1.2 \, R_{\rm {inf}}$ here. Figure adapted from \citet{nakamura13}; reproduced by permission of the AAS.}
\label{fig:m87rR}
\end{figure}

The high frequency, high resolution VLBI observations of M87 show the jet accelerating smoothly, with a parabolic profile, to relativistic speed over a deprojected radius $\sim300\,{\rm pc}\sim R_{\rm inf}$ \citep{nakamura13, walker18}.  At this point a strong, quasi-stationary shock, HST-1, appears to form in the flow (\textbf{Figure~\ref{fig:m87mon}}) and regularly ejects superluminal components \citep{cheung07, nakamura10} at apparent speeds up to  $6\, {\rm c}$ \citep{biretta99, gir12} before the jet slows and widens (\textbf{Figure~\ref{fig:m87rR}}).  The helical structure observed beyond HST-1 suggests that the magnetic field is still dynamically strong here \citep{asada12}. A similar, stationary transition is observed near the component C7 in BL Lac. This is also at a deprojected radius $\sim R_{\rm inf}$ and also creates superluminal components.

\paragraph{3C84 Imaging}\label{par:3c84}Recent observations of 3C84 tell a somewhat different story \citep{hodgson18}. The jet appears to expand to create a strongly edge-brightened cylinder with a diameter $\sim1000\,r_{\rm g}$, increasing slowly with height, out to a de-projected distance $\sim3\times10^4 \, r_{\rm g}$, well within $R_{\rm inf}$. The jet appears to be propagating through and interacting with a constant pressure cocoon, either a very hot gas or, more plausibly, a magnetized sheath.  

\paragraph{Bl Lac. Imaging}\label{par:bllac}Very high resolution VLBI imaging of BL Lac indicates that the structure of the ``unresolved VLBI core'' is actually very rich and complex. MOJAVE observations of BL Lac \citet{coh14} show that a stationary component (C7, probably a stationary shock) lies 0.26 mas from the true core (where the black hole is believed to reside) or $\sim 3 {\rm pc}$ deprojected, assuming a BL Lac viewing angle of $\sim 6\degr$. The stationary component appears to be the source of the ejected superluminally-moving components seen in BL Lac outbursts. Furthermore, \citet{coh15} find that, in addition to these moving components (identified as longitudinal fast waves or shocks in the jet's helical magnetic field moving at a speed $\beta_{\rm F} \sim 0.68$ in the rest frame of the $\Gamma \sim 4.5$ jet) there are transverse Alfv\'en waves traveling downstream at a slower velocity ($\beta_{\rm T} \sim 0.25$). These results, along with the generally observed longitudinal EVPA in BLL \citep{gmc04}, are further evidence that these jets are dominated by a strong helical magnetic field even downstream of the C7 shock.

\paragraph{Jet Collimation}\label{par:jcb}In M87, unlike BL Lac, the HST-1 region has not been examined for long-term traveling transverse Alfv\'en waves, but there was a transverse shift in the position angle of the HST-1 ejecta from $-65 \degr$ to $-100 \degr$ over the period 2007.00--2011.65. Interestingly, if the $\sim 10^{9.8} \, {\rm M}_{\odot}$ M87 black hole system \citep{wbhs13} is simply scaled linearly down to the size of the $\sim 10^{7.5} \, {\rm M}_{\odot}$ BLL \citep{ts17}, the distance between HST-1 and the M87 black hole shrinks from $\sim 300$ pc to $\sim 3$ pc, similar to the distance between C7 and the BL Lac black hole. 

Studies of the M87 jet width upstream and downstream of HST-1 have revealed important clues to jet acceleration, collimation, and propagation \citep{nakamura13}. Downstream of HST-1 (the analog of the ``jet'' in traditional VLBI core-jet sources) the flow has a fixed conical opening angle. This result justifies the assumption of a relativistic conical flow in theoretical work on VLBI jets \citep{blandford79}. Upstream of HST-1, however, the jet shape is semi-parabolic, with a profile of radius $r \propto$ SMBH distance $R^{0.58 \pm 0.02}$, which requires a modification of this model \citep{algaba17}. Astonishingly, even higher resolution observations with the VLBA up to 86 GHz and with the EHT at 230 GHz \citep{doeleman12} already show that this profile persists all the way down to $R\sim6r_{\rm g}$ above the black hole. 

\paragraph{Polarization and Acceleration}\label{par:acz}Detailed studies of the linear polarization in M87 at radio and optical wavelengths,  have been carried out, with the VLA and HST, as described by \citet{avachat15}.  The most striking findings are that (i) both images show similar polarization structure near the nucleus; (ii) with the magnetic vectors parallel to the jet axis; (iii) at the knot HST-1 the  degree of polarization is significantly greater in the optical images and the magnetic vectors are perpendicular to the jet axis.  This is interpreted as as showing that the higher energy emission along the jet comes from deeper within the jet and that the shock at HST-1 is deeply embedded in the jet.

Finally, proper motion studies of the M87 jet \citep{andni14} show that acceleration is steady from an apparent speed of $\sim 0.01 \, {\rm c}$ at a black hole distance of $\sim 400 \, {\rm r}_g$ to a superluminal peak of $\sim 6 \, {\rm c}$ at HST-1 ($\sim 5 \times 10^5 \, {\rm r}_g$), and then deceleration occurs to $\sim 0.4 \, {\rm c}$ at $2 \times 10^7 \, {\rm r}_g$ from the black hole.

To summarize, in low-luminosity FR-I sources, jet acceleration and collimation to superluminal speeds appears to occur within the ``unresolved VLBI core'' over a distance of $\sim 10^{5-6} \, {\rm r}_g$. This ACZ terminates in a strong shock, which then ejects the multiple superluminal jet events that have been seen in VLBI observations since the early 1980s. The high speed of relativistic jets, therefore, does not develop within a short distance from the black hole, but rather is produced slowly over a distance of hundreds of thousands of black hole radii as the jet propagates out of the galactic center. 

\subsubsection{Lobe Jets ($0.1R_{\rm lobe}\lesssim R<R_{\rm lobe}$)}\label{sssec:lobes}
\paragraph{FR Class}\label{par:frc}Radio maps of AGN made with the Cambridge One Mile Telescope showed that many radio sources are double-lobed like Cygnus A and that they have fairly compact regions, or hot spots, in their lobes as well as the compact cores in their nuclei.
\begin{marginnote}[2pt]\entry{FR}{Fanaroff-Riley Class}\end{marginnote}
This supported continuous energy supply from the nucleus to the lobes by jets \citep{hargrave74}. A major advance in classification was the division of double radio sources into lower radio luminosity, FR-I sources, in which the highest brightness regions are closer to the nucleus than the low brightness extremities, and higher radio luminosity FR-II sources when they are not \citep{fanaroff74}.

\paragraph{FR-II Jets}\label{par:friijet}
FR-II jets are often quite asymmetric and the brighter one is accompanied by lower Faraday rotation verifying that it is approaching and is still Doppler-beamed \citep{garrington88}. So, FR-II jets appear to emerge from their galactic nuclei with, at least, mildly relativistic speed and a significant fraction of their initial power. The first FR-II jet was discovered in 3C219 \citep{turland75} and this was soon followed by a spectacular 200 kpc-long jet in NGC6251 \citep{waggett77}.  

\begin{figure}[ht]
\includegraphics[width=5in]{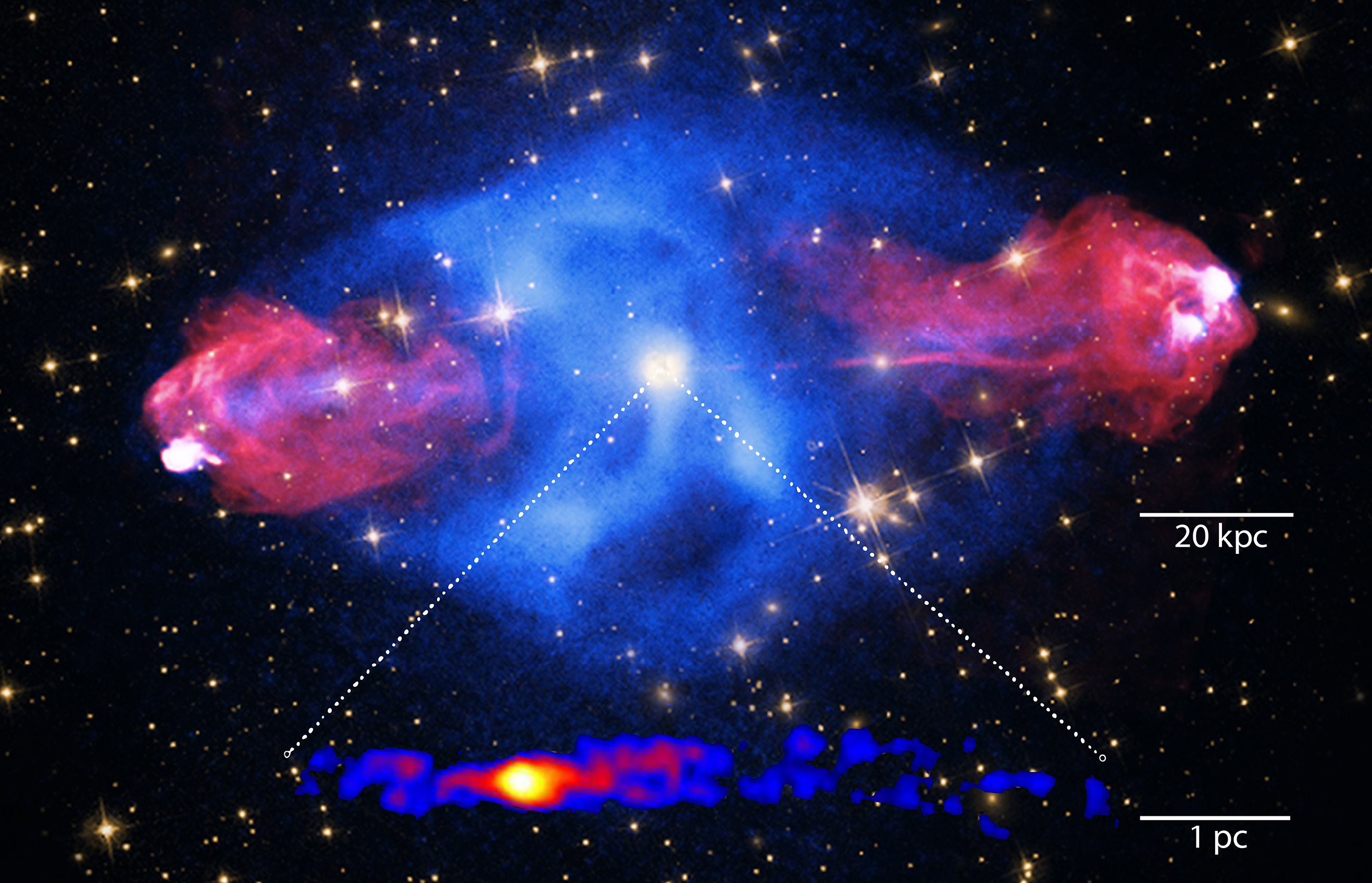}
\caption{Comparison of radio (red), X-ray (blue), and optical images of the Cygnus A FR-II radio galaxy (center). While the X-ray traces primarily hot old cocoon emission and the radio traces the jet and new cocoon emission, the three primary hot spots are bright in both energy bands with a synchrotron peak in the radio and a Compton scattering peak in the X-ray. The nearly-symmetric black hole jet is shown at bottom, with a scale 20,000 times smaller than the larger image. Credit: X-ray: NASA/CXC/SAO; Optical: NASA/STScI; Radio: NSF/NRAO/AUI/VLA; VLBI inset: \citet{boccardi17}.}
\label{fig:cyga}
\end{figure}

\paragraph{FR-I Jets}\label{par:frijet}Much has been learned from a systematic VLA polarimetric survey of ten FR-I jets by \citet{laing14}. It was concluded that they expand faster than linearly within  1-30~kpc -- after which they expand linearly. Meanwhile, the magnetic field transitions from axial to toroidal. A slower jet boundary later is observed and and a velocity gradient can be inferred across the jet. Spectral data, including X-ray observations, imply that particles are being continuously accelerated in the jet. 

\paragraph{CSO Jets}\label{par:cso}CSOs, \citep{wilkinson94} are very likely young FR-II objects \citet{readhead96}.  
\begin{marginnote}[2pt]\entry{CSO}{Compact Symmetric Object}\end{marginnote}
These have been shown to indeed be young and to have ages in the range 20-2000 years \citep{gugliucci05}. The first large uniform sample of CSOs to be studied is that of \citet{tremblay16} who show that the CSO class exhibits both FR-I and FR-II morphology, and, furthermore, the morphology appears to depend on luminosity, with the transition occurring at the same luminosity as in their larger cousins.

\subsection{$\gamma$-ray Observations to Very High Energies (VHE)}\label{ssec:gam}
The launch of  Fermi in 2008 and its highly successful operation over the last decade has had a tremendous impact on studies of blazars.  Fermi surveys the whole sky every 3 hours over the energy range 20 MeV -- 300 GeV. There have been corresponding advances in ground-based atmospheric- and water-Cerenkov telescopes, including HAWC, HESS, MAGIC and VERITAS, that have extended the spectral range to energies beyond $\sim10\,{\rm TeV}$ (and can have sensitivity to $\sim 100\,{\rm TeV}$) in some AGN \citep{funk15,lott15,lindfors15,gonzalez15,lauer15}. $\gamma-$ray observations complement radio studies with an extensive --- up to 20 octave --- SED
\begin{marginnote}[2pt]\entry{SED}{Spectral Energy Distribution}\end{marginnote}
and rapid variability. These remarkable advances in $\gamma-$ray capability have spearheaded a large global program of multi-wavelength and multi-messenger observations. 

In a radio flux density-limited sample of flat-spectrum AGN selected at 5 GHz it was found that only $\sim20\%$ of the objects are detected in $\gamma-$rays \citep{karouzos11}. However all of the Fermi-detected blazars north of declination $-20^\circ$ are detected in the OVRO 40 m Telescope monitoring campaign \citet{richards11}. \citet{karouzos11} find no strong link between fast apparent speeds and $\gamma-$ray detectability, as measured with Fermi. They argue that this is evidence for a ``spine-sheath'' structure \citep{sol89,laing96} in which the outer layers of the relativistic jet, which form  the ``sheath'' have slower bulk velocity along the jet axis than do the inner layers, which form the ``spine''. Such a structure could also explain the relative fractions of AGN that are $\gamma-$ray bright and radio bright,  if the spine is predominantly  $\gamma-$ray-emitting and the sheath is predominantly radio-emitting. 
 
\begin{figure}[ht]
\includegraphics[width=4in]{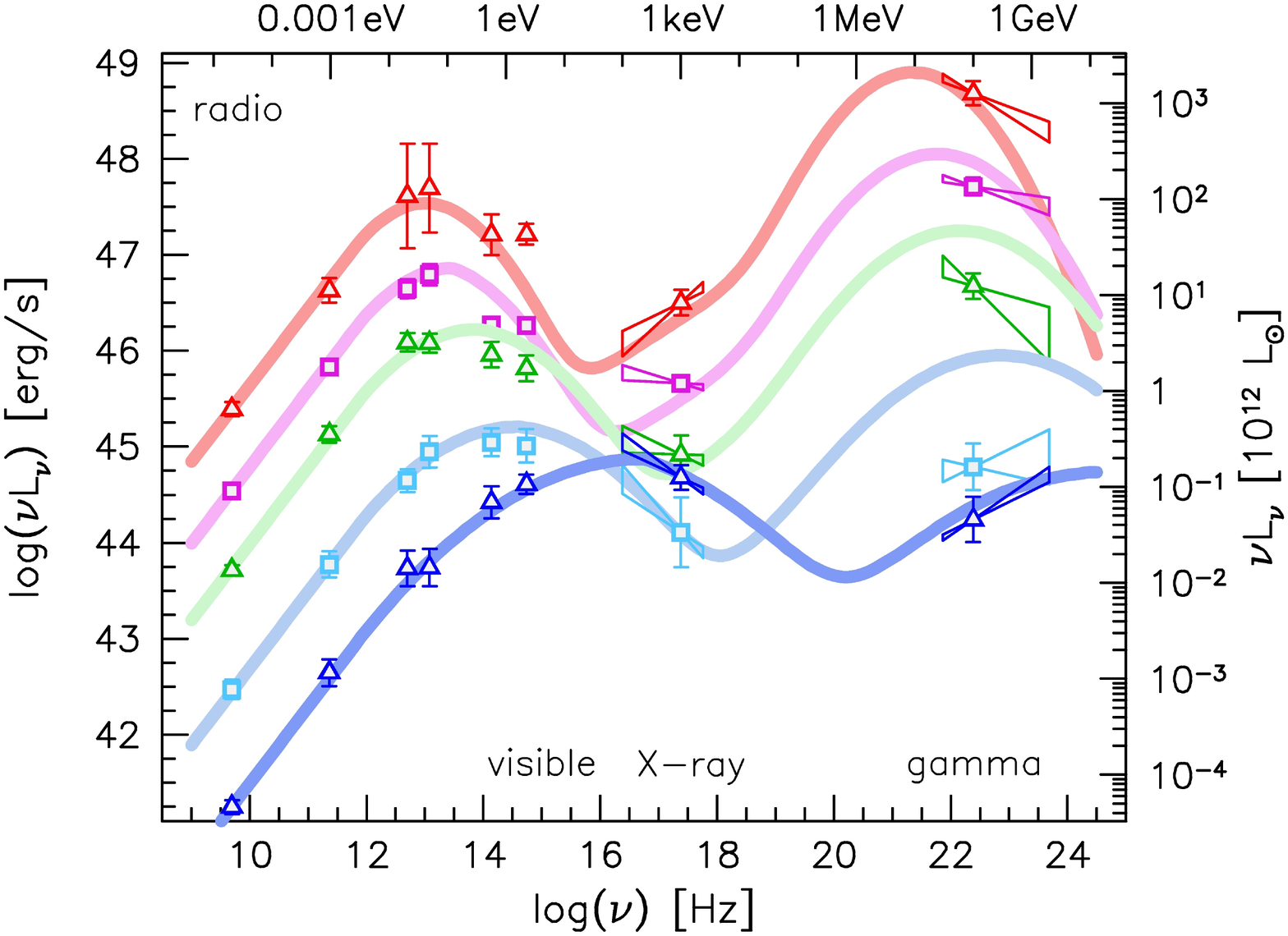}
\caption{The Blazar Sequence (\textbf{Table~\ref{tab:BLSEQ}}) in the electromagnetic spectrum, showing a high-powered FSRQ (red), LBL (pink), medium-luminosity IBL (green), HBL (light blue), and low-powered extreme HBL/TeV (dark blue) blazar. After \citet{fossati98}}
\label{fig:blseq}
\end{figure}

In a comprehensive review of $\gamma-$ray observations of AGN, \citet{madejski16} found that the $\gamma-$ray flux density variations in blazars show generally greater fractional amplitudes than the other observed bands, and stronger flares tend to occur when the flux density is at a higher-than-average level, with the activity lasting anywhere from several days to several months. It is useful to consider two subclasses of AGN based on the Blazar Sequence of \citet{ghisellini98, ghisellini16}: (i) the high-synchrotron-peaked BLL (HBL), low luminosity,  line-less BLL; and (ii) the powerful flat spectrum radio quasars (FSRQs), which have high luminosities and strong emission lines. (See also \textbf{Tables~\ref{tab:RLAGN}} \& \textbf{\ref{tab:BLSEQ}} and \textbf{Figure~\ref{fig:blseq}}.) The FSRQ-type blazars show greater amplitude of $\gamma-$ray variability than the HBL-type blazars, but HBL blazars show the greatest variability amplitude in the VHE $\gamma-$rays. Furthermore the FSRQs have soft $\gamma-$ray spectra, while the HBL blazars have hard $\gamma-$ray spectra. \citet{madejski16} conclude that the broadband data strongly suggest that the primary dissipation of the jet energy converted to blazar power at $\gamma-$ray energies occurs beyond $\sim10^4r_{\rm g}$ and that the size of the dissipation region as inferred from the $\gamma-$ray variability timescale is very small.

\begin{table}[ht]
\tabcolsep7.5pt
\caption{Classification of Radio Loud AGN by their Radio Properties}
\begin{center}
\begin{tabular}{@{}l|c|c|c@{}}
\hline
Abbrev.&Expansion&Definition&Density ($L_{\rm bol}$)$^{\rm a}$\\
\hline\hline
FR-I & Fanaroff-Riley type 1 & Low power, & $\sim10^4$ ($<10^{42}$)\\& & edge-darkened radio source  & \\
\hline
FR-II & Fanaroff-Riley type 2 & High power, & $\sim10^{1.5}$ ($>10^{42}$)\\& & edge-brightened radio source & \\
\hline\hline
BLL & BL Lac object & Compact radio source with & $\sim10^{2.3}$ ($<10^{42}$)\\& & polarized optical continuum & \\& & and weak or no emission lines & \\
\hline
FSRQ & Flat Spectrum & Compact radio source & $\sim10^{0.8}$ ($>10^{42}$)\\& Radio Quasar & identified with a quasar & \\
\hline
\end{tabular}
\end{center}
\begin{tabnote}
$^{\rm a}$ Gpc$^{-3}$, local; AGN $L_{\rm bol}$ in erg $s^{-1}$, \citet{tadhunter16,braun12,ajello14} 
\end{tabnote}
\label{tab:RLAGN}
\end{table}

A corresponding review of VHE observations of AGN \citep{prandini17}, shows that 70 AGN have  been detected, including 3C279 (z=0.54), PKS 1441+25 (z=0.94), and S3 0218+35 (z=0.95). The highlights are as follows: (i) most of the TeV AGN are HBLs; (ii) the only extragalactic non-AGN sources are the nearby starburst galaxies M 82 and NGC253; (iii) the large majority of sources are at $z<0.5$ and all are at $z<1$; (iv) a number of detected objects have no redshift measurement, as is not atypical of BLL; (vi) although FSRQs form the majority of AGN detected by Fermi, they make up only a small fraction of the TeV-detected sources, as might be expected from the blazar sequence; (vii) seven FSRQs have been detected at the highest energies, with two of them being the highest redshift AGNs detected at VHE thus far,  and one of them being the gravitationally lensed system B0218+357; and (viii) the high redshift systems place important constraints on our estimates of the extragalactic background light (EBL).

\subsection{OIR, UV and X-ray Observations}\label{ssec:oiru}
HST, Chandra and NuSTAR, especially, have also transformed our understanding of jets by connecting the radio to the  $\gamma$-ray emission and helping us understand where particles are accelerated along jets and to what energies. X-rays in AGN originate from the AGN accretion disk itself as well as from the jet.  The general properties have been reviewed by \citet{brandt15}, and a detailed discussion of the X-ray variability in 81 AGN observed with NuSTAR is given by \citet{rani17}. 65\% of their sources show significant variability on hourly timescales. There are several comprehensive reviews of X-ray observations of relativistic jets \citep{harris06,sambruna12,schwartz15}.  A key discovery of Chandra has been that of the emission from almost 100 relativistic jets, some of which extend up to hundreds of kpc (see, e.g., https://hea-www.harvard.edu/XJET/).   The X-ray jets detected by Chandra are dominated by bright knots of emission. In the powerful objects they also terminate in hot spots (\textbf{Figure~\ref{fig:cyga}}). It has been found that powerful FR-II jets are 100x too bright to be radiating by Synchrotron Self Compton emission.  The possibility that the electrons are scattering CMB photons has been suggested, but this requires high bulk Lorentz factors and angles close to the line of sight. A comprehensive discussion of this problem and its possible solutions has been given by \citet{georganopoulos16}.

A prime example is that of the jet in 3C111, which \citet{clautice16} have studied with both Chandra and the HST.  As is typical in these cases, the jet is seen in X-rays through knots that coincide with radio knots and in the hot spots in which the jets terminate. The combined observations and implied velocity of the approaching hot spot strongly disfavor EC/CMB models in favor of a two-component synchrotron model, thereby providing strong evidence for the multi-zone model.

The jet in M87 has been studied intensively with Chandra \citep{marshall02}. Almost all the optically bright knots are seen in X-rays, with the core being the brightest  feature. Synchrotron models fit the knots in the jet well.

The combined X-ray results from Chandra, Suzaku, and Swift observations of 3C273 \citep{madsen15}, and especially the high energy NuSTAR, multi-epoch INTEGRAL, and Fermi observations, make it possible to separate out the coronal emission plus disk reflection component from the jet component. They show that the jet component can be fitted by a log-parabola model that peaks at $\sim 2$ MeV.

\begin{figure}[ht]
\includegraphics[width=4in]{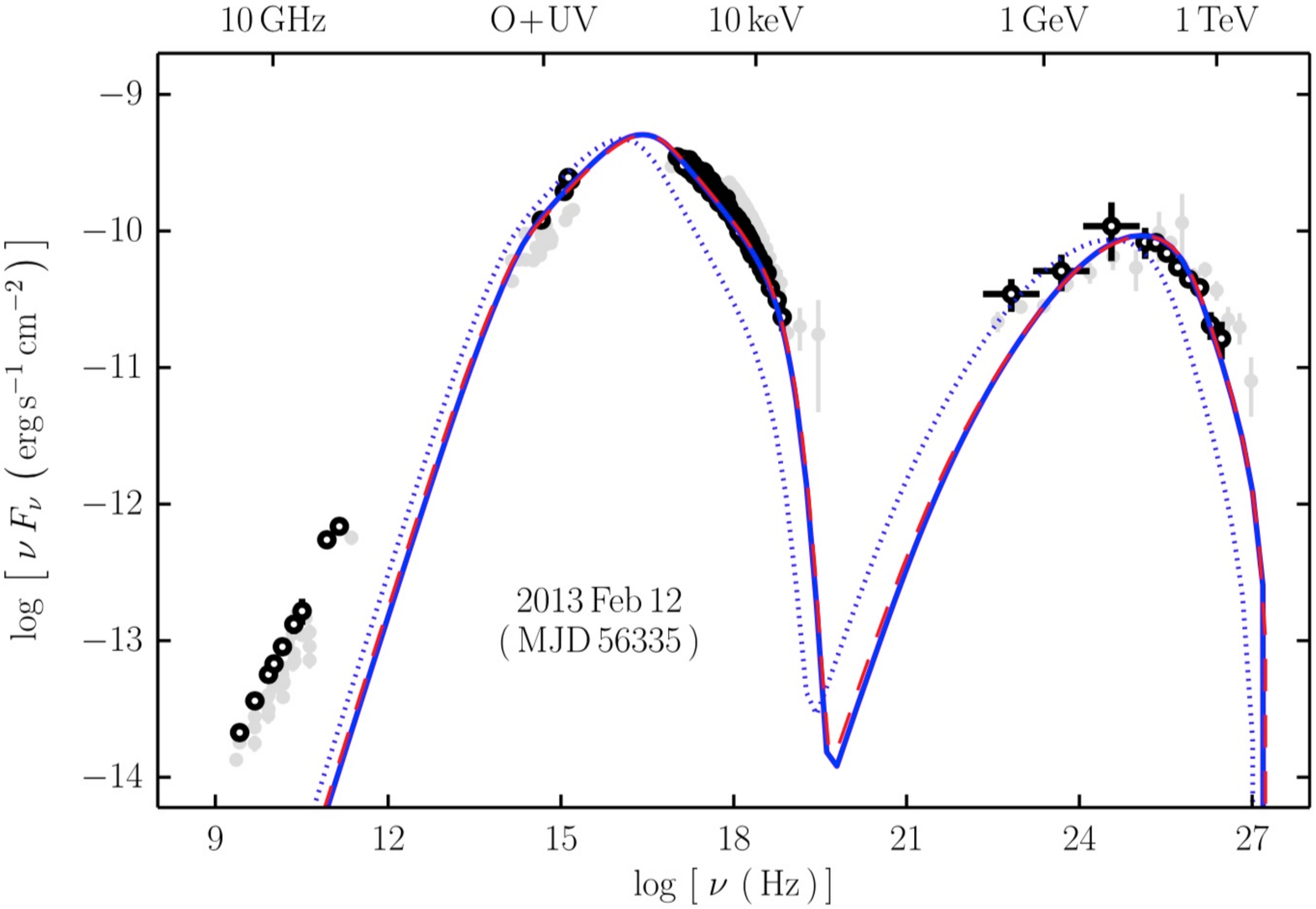}
\caption{SED of Mkn421 combining observations from GASP-WEBT, Swift, NuSTAR, Fermi, MAGIC, VERITAS, and other and instruments \citep{balokovic16}. The solid blue lines assume a simple one-zone SSC model, which these authors argue against based on the fact that the synchrotron cooling time is a factor 30 shorter that the variability timescale. The grey symbols in the background are from \citet{abdo11}.}
\label{fig:bactrian}
\end{figure}

An example of the results of a well-coordinated multi-band campaign to determine the  SED of  an AGN (in this case Mkn421), combining observations from GASP-WEBT, Swift, NuSTAR, Fermi, MAGIC, VERITAS, and other and instruments, is shown in (\textbf{Figure~\ref{fig:bactrian}}), which is taken from  \citet{balokovic16}. The ``Bactrian'' shape of the SED is typical of AGN, and it is by now well-established that the lower-frequency peak is due to synchrotron emission. Although the higher-frequency peak is widely believed to be due to inverse Compton scattering, as we have seen above it is possible that it too might at least in some cases be due to synchrotron emission. Observations of an individual $\gamma$-radio flare in M87 support the view that the $\gamma$-rays originate within $\sim100 \, r_{\rm g}$ \citep{hada14}.
 
Many SED studies of AGN use the ``one zone model'', which assumes that the emission causing both the synchrotron peak and the inverse-Compton peak originates from the same particle population. This assumption has been challenged in view of the  degree of complexity seen in the radio emission regions, which probably persists down to the smallest scales in these extremely energetic objects. In the case of Mkn421 \citet{balokovic16} show that the synchrotron cooling timescale $\sim 1000$ s, whereas $\tau_{\rm var} \approx 9$ hr, showing that it is unlikely that the emission is dominated by a single shocked region.  This argues for a multi-zone model.

\subsection{UHE Cosmic Rays and VHE Neutrinos}\label{ssec:crnu}
Relativistic jets associated with SMBH are plausible sites for the production of both High Energy (PeV-EeV) and Ultra High Energy Cosmic Rays ($\sim1-200\,{\rm EeV}$) \citep{matthews18}. This is because the potential difference across across an electromagnetic jet powered by a spinning black hole and carrying a power $L_{\rm jet}\sim10^{45}\,{\rm erg\,s}^{-1}$ is $\sim100\,{\rm EV}$ (\textbf{Sidebar Electromagnetic Effects}). (It is unknown how much of this potential is actually available for particle acceleration, but it is worth noting that the Crab pulsar which has a rotational EMF of $\sim50\,{\rm PV}$, manages to accelerate electrons in the nebula to at least $\sim3\,{\rm PeV}$.)  There are only a few other sites where the highest energy particles could be created \citep{hillas84}, \textbf{Section~\ref{sssec:paccnP}}. A concern with this proposal is that, although high energy hadrons may be accelerated in AGN jets, they may also be subject to catastrophic losses (\textbf{Section~\ref{par:pgpi}}).

Neutrinos with energies in the TeV to a few PeV range have been detected and associated with AGN jets. Recently, the possible identification of a single neutrino emitted with an energy $\sim400\,{\rm TeV}$ with a flaring blazar, TXS 0506+056, has been reported \citep{aartsen18}. However, the lack of a correlation of the arrival directions of other neutrinos with known blazars suggest that no more than a quarter of these neutrinos derive from Fermi blazars \citep{aartsen17}.

\begin{textbox}[ht]
\section{ELECTROMAGNETIC EFFECTS}
These simple accretion disks conserve mass and angular momentum. Energy is transported radially outward and the inevitable energy dissipation is supposed to be balanced by radiative loss from the disk surfaces. However, in the absence of radiative loss, a MHD wind can remove angular momentum and energy in the same ratio as they are released by the inflowing matter, obviating the need for dissipation \citep{blandford76,lovelace76,blandford82}. The power and mass loss in the wind depend upon subtle physics, just as is the case with the solar wind. If there is open magnetic flux $\Phi_{\rm disk}$ threading a portion of the disk orbiting with angular velocity $\Omega_{\rm disk}$ then the wind power may be estimated (in S.I. units) by $L_{\rm disk}\sim\Phi_{\rm disk}^2\Omega_{\rm disk}^2/\mu_0V_{\rm A\,crit}$, where $V_{\rm A\,crit}$ is the Alfv\'en speed at the critical point.

In addition, a portion $[1-(r_+/2r_{\rm g})^{1/2}]Mc^2$ of the mass of the hole, is associated with its spin \citep{penrose69} and is extractable by electromagnetic stress \citep{blandford77,mckinney07,penna13}. The power extracted can be calculated by solving the Einstein-Maxwell equations assuming an external current distribution and regularity at the horizon. The induced EMF is $V\sim\Omega_{\rm H}\Phi_{\rm H}/2,$, where $\Phi_{\rm H}$ is the magnetic flux threading the horizon. The effective resistances associated with the horizon and a relativistic outflow are both $\sim50\,{\rm Ohm}$ and so the jet power and current are $L_{\rm jet}\sim10^{45}(V/100\,{\rm EV})\,{\rm erg\,s}^{-1}$, $I\sim(V/100\,{\rm EV})\,{\rm EA}$, respectively. This hole will gain mass at a rate $\sim5\times10^7\,(L_{\rm jet}/10^{45}\,{\rm erg\,s}^{-1})\,{\rm M}_\odot\,{\rm Gyr}^{-1}$ while losing spin energy at twice this rate.
\end{textbox}

\subsection{Variability: Relative Locations of the Multi-band Emission Regions}\label{ssec:var}
A key question in AGN studies is that of the relative locations of the emission regions in different bands.  The earliest VLBI images showed a clear pattern in which the emission from higher frequencies occurs closer to the central engine \citep{readhead78b}. Given the one-sided jet morphology \citep{wrpa77} it was immediately clear  that higher frequency observations were penetrating the optically thick, flat-spectrum, core at the end of the jet and probing deeper into the jet towards the SMBH. This pattern has now been found to be the case in most blazars observed with VLBI. In general, therefore, at radio frequencies the higher frequency emission is located closer to the SMBH than the lower frequency emission.  The interesting question here is whether the same is true when going to even higher energy bands, including $\gamma-$rays. Since only VLBI offers milli-arcsecond resolution, for the most part the study of the relative locations of the emission regions in different observing bands has depended on variability and SED studies. 

\begin{table}[hb]
\tabcolsep7.5pt
\caption{The Blazar Sequence}
\begin{center}
\begin{tabular}{@{}l|c|c|c@{}}
\hline
Abbrev. & Expansion & Probable  Radio Parent & Emission Lines\\
\hline
\hline
Extreme HBL & TeV Blazars (BLL) & Low-luminosity FR-Is & Weak\\
\hline
HBL & High-energy peaked (blue) BLL & FR-I Sources & Weak\\
\hline
IBL & Intermediate-energy peaked BLL & FR-I/II Break Sources& Weak\\
\hline
LBL & Low-energy peaked (red) BLL & Class B FR-IIs & Weak\\
\hline
FSRQ & Flat Spectrum Radio Quasar & BLRG, FR-II QSR & Strong\\
\hline
\end{tabular}
\end{center}
\label{tab:BLSEQ}
\end{table}

There have been many radio monitoring surveys over the last five decades.  Here we focus on three of the more recent large-scale surveys that are particularly important for the determination of the relative locations of the $\gamma-$ray and radio emission regions in AGN: (i) the decades long University of Michigan Radio Astronomy Observatory (UMRAO) blazar polarization survey at 4.8 GHz, 8.0 GHz,  and 14.5 GHz \citep{aller17}; (ii) the Caltech OVRO 40 m Telescope monitoring survey of 1158 AGN at 15 GHz \citep{richards11, moerbeck14}, and (iii) the Max Planck Institute for Radio Astronomy (MPIfR) multi-frequency ``FGAMMA'' monitoring  survey of 54 Fermi-bright blazars \citep{fuhrmann14}. The UMRAO program showed that the linear polarization is generally at the level of a few percent during  a flare but can reach as  high as 15\% in some sources at 14.5 GHz. In the Caltech OVRO 40 m program the BLL showed larger variability amplitudes than flat spectrum radio quasars (FSRQs). The FGAMMA observations show clear migration of flares from high frequencies to low frequencies with time.   A recent review of intrinsic variability and interstellar scintillations (ISS) is that of
\citet{jauncey16}.  Intra-day variations due to ISS are seen at the  2\% - 10\% level in $\sim50$\% of flat-spectrum radio sources at cm wavelengths. A larger fraction of the weaker sources show ISS.

 Both the Caltech 40 m and the MPIfR FGAMMA studies found statistically significant correlations between the radio- and $\gamma-$ray flux density variations in some of the AGN studied \citep{richards11, moerbeck14, fuhrmann14}. In all such cases the cm wavelength radio variations lagged behind the $\gamma-$ray variations by about a hundred days.  The FGAMMA results also showed that the lag decreases with wavelength, falling to zero by $\approx100$ GHz. It is therefore clear that in some AGN the $\gamma-$ ray emission regions are indeed closer to the SMBH than the radio emission regions.  However there are also observations that show that in some AGN the $\gamma-$ray and radio emission regions are co-located \citep{jorstad01, marscher11, jorstad16, rani18}.  It appears therefore that $\gamma-$ray emission from AGN occurs at a variety of sites, some close to the SMBH and some close to the radio-emitting regions. \citet{nalewajko14} have presented an observational plan to locate the sites of synchrotron and Compton emission and argued that spine-sheath jet structure models are not a plausible alternative to external broad emission line photons.

\begin{figure}[ht]
\includegraphics[width=4in]{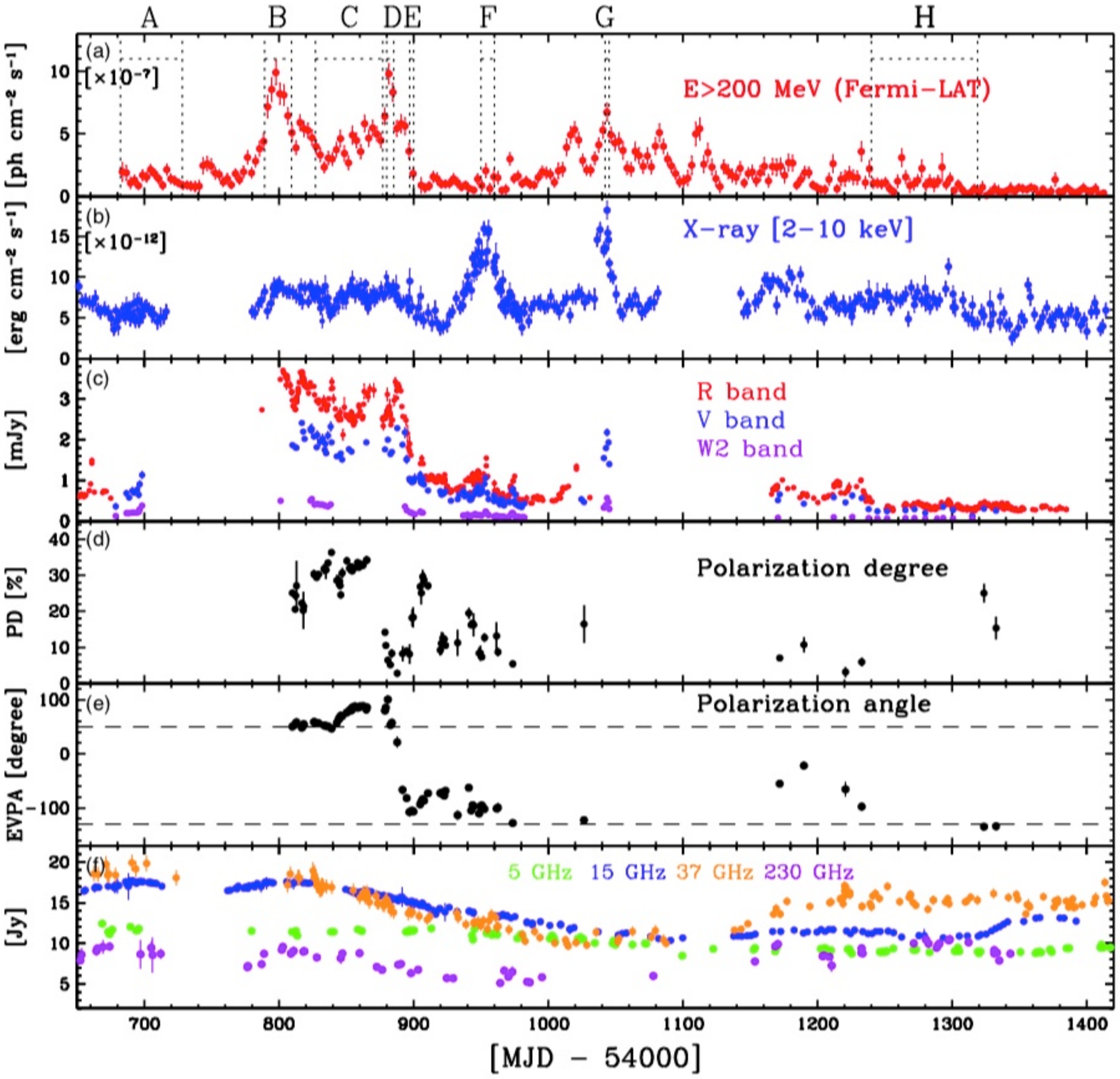}
\caption{Multi-band light curve of 3C279 from August 2008 to August 2010 from \citet{hayashida12}, illustrating the complex nature of the variations in the different bands and the challenge in associating variability in one band with that in another given that more than one zone must often be responsible for the radiation in even a single band and that there can be different light travel times from these zones.}
\label{fig:3c279}
\end{figure}

It is instructive to consider the variability in the major energy bands of the archetypal blazar 3C279 shown in \textbf{Figure~\ref{fig:3c279}}, which is taken from \citet{hayashida12}. We note that in observing single objects for relatively short periods of a few years it is very difficult to estimate the significance of any apparent correlations.  Here the emission is discussed in terms of the apparent correlation of variability in the $\gamma-$ray and optical bands, and the lack of correlation with the X-ray band using the one zone model and it is shown that the variability can be modeled by optical and $\gamma-$ray emission components located in the radio emitting components at distances of several pc from the SMBH.  However, it is also shown that the observations are consistent with $\gamma-$ray/optical emission that is generated much closer to the SMBH. 

\subsection{Other Jets}\label{ssec:other}
There are many other examples of the jet phenomenon. The most similar are the ``Galactic Superluminal Sources'', produced by stellar mass black holes in binary sources \citep{mirabel99}, SS433 \citep{abeysekara18}, Gamma Ray Bursts \citep{gehrels09}, (including GW170817\citep{mooley18}), Pulsar Wind Nebulae \citep{durant13} and Protostellar Jets \citep{bally16}. Although the physical conditions are quite different in each of these classes of source, common principles are invoked in attempts to model them. 

\section{JET KINEMATICS AND DYNAMICS}\label{sec:kindyn}
\begin{extract}
``\dots If, for example, $\gamma=5$, the apparent diameter of the source will increase by almost 10 light years each year\dots an expanding source could exhibit a rate of increase of flux density high enough to explain the observations.''-- \citet{rees66}
\end{extract}

\subsection{Dynamics and Energetics of Extended Sources}\label{ssec:dynex}

\subsubsection{Energy Generation in the Galactic Nucleus: Black Holes and Accretion Disks}\label{sssec:engen}
It is generally believed that a SMBH exists at the center of nearly all large galaxies with a significant central stellar bulge \citep{kormendy13}. In all such galactic nuclei gas eventually, through various means, will make its way from the outer galaxy to the vicinity of the SMBH. The gas can come from a merger with a gas-rich galaxy, the galaxy's own interstellar medium (ISM), gas shed by red giants, supergiants, and supernnovae, and the tidal disruption of stars that venture inside the SMBH's tidal radius. The gas will have an angular momentum about an initial axis determined by parameters external to the SMBH. However, if the SMBH is spinning rapidly, the gas will tend to orient in the latter's equatorial plane near the hole, forming an accretion disk. When in a steady state, the rate of matter inflow toward the SMBH in this disk is given by the parameter $\dot{M}$ in units of mass per unit time, which is often scaled to the Eddington accretion rate (\textbf{Sidebar Black Hole Accretion}). It is generally convenient to describe the state of the plasma using plasma $\beta$ parameters which normalize the pressure of various components to the magnetic pressure.

 There are two ways for a jet to be formed by a spinning, accreting, black hole: the disk and the spinning black hole itself (\textbf{Sidebar Electromagnetic Effects}). If a \citet{shakura73} type disk surrounds the hole, the disk and hole jet luminosities are given by $L_d \sim 0.5 \, L_{Edd} \, \dot{m}^2$ and $L_h \sim 0.5 \, L_{Edd} \, \dot{m}^2 j^2$ \citep{meier12} for near-Eddington accretion rates. If a black hole in the range $m \sim 10^{8-10} \, {\rm M}_{\odot}$ is spinning rapidly, and its accretion rate is, say, half the Eddington limit, in principle both these mechanisms are capable of generating jet powers of $\sim 10^{45-47} \, {\rm erg \, s^{-1}}$. Over a lifetime of, say, $10^7$ yr such a jet can pump up to $\sim 10^{59-61}$ erg into the radio lobes. 

\subsubsection{Possible Accretion States in AGN Black Holes}\label{sssec:accstate}
While we do not discuss jets in black hole systems with lower mass ({\textit e.g.}, X-ray binary systems) in any detail here, it is very instructive to review accretion states in these systems (which can be observed over many accretion time scales) and how these states relate to jet production in these systems. Black holes of $3 - 30 \, {\rm M}_{\odot}$ are observed to have {\em four} main accretion states. At the lowest accretion rates, the disk exists in the low/hard (L/H) state, where it is very geometrically thick and optically thin; in this state, the BH system is frequently associated with a weak, steady jet. At higher accretion rates, the disk enters the high/soft (H/S) state, where the accretion inflow is geometrically thin and optically thick; jets are rarely, if ever, seen in the H/S state. When the accretion rate approaches the Eddington value (\textbf{Sidebar Black Hole Accretion}), the very high/unstable (VH/U) state is seen, in which the inflow displays complicated time-dependent behavior, sometimes entering a soft sub-state for a long time and then briefly entering a hard sub-state and producing a jet. Finally, at accretion rates significantly above the Eddington value, the accretion inflow produces a strong, optically thick, radiation-pressure-driven wind that can enshroud much of the system and look like a supersoft X-ray source. 

Detailed, time-dependent models and simulations of these four states, and transitions between them, are not yet available. But if one can appeal to simple scaling models, one can suggest how they relate to accretion disk theory:  the L/H state occurs for $\dot{m} < \dot{m}_A$, the H/S for $\dot{m}_A < \dot{m} < \dot{m}_I$, the VH/U state for $\dot{m}_I < \dot{m} < 1$, and the super-Eddington state for $\dot{m} > 1$ (\textbf{Sidebar Black Hole Accretion}). AGN should then have only three accretion states:  L/H, VH/U, and super-Eddington. In these models, the truly radio silent H/S state disappears because of the mass dependence of $\dot{m}_I$. Low-luminosity AGN (LLAGN) and FR-I sources then would be expected to be in the L/H state and produce a weak steady radio jet. High-luminosity AGN (Seyferts and quasars) then may be examples of the VH/U state, usually in a soft sub-state, but entering a hard state for a relatively short time and producing a radio jet, on a cycle time of potentially hundreds to millions of years \citep{meier12}. Detailed simulations of accretion flows around black holes over a wide range of mass and accretion rate, plus observations of intermediate mass black hole (IMBH, $100 < m < 10^5$) accretion state behavior, could shed light on these somewhat preliminary concepts for AGN. 

\subsubsection{Energy Dissipation in the Jet and Extended Lobes}\label{sssec:endiss}
Energy dissipation via particle acceleration and radiative emission is covered in detail in \textbf{Section~\ref{sec:emmod}}.  We discuss only a few key points here to complete the process from energy generation at the black hole, transport of that energy to regions at distances ranging from mpc to Mpc and finally dissipation of that energy at those sites.

Observations discussed in \textbf{Section~\ref{sec:obs}} indicate that in all sources energy is possibly dissipated all along the jet, from black hole to lobes. However, in many FR-I jets a great deal of dissipation apparently occurs at a probable recollimation shock near $R_{\rm inf}$.  Astoundingly, however, even though the jet is often slowed there to moderately-relativistic speeds, it continues to carry energy to sites that are orders of magnitude further in distance, especially in electromagnetic and particle forms.  This is evidenced by the large amount of emission in the Galaxy and lobe jets and the lobes themselves (\textbf{Figure~\ref{fig:m87mon}}). 

FR-II jets are equally mysterious.  While their black hole jets are certainly accelerated and collimated by electromagnetic means, by the time the lobe jets reach the hot spots and lobes, they behave and emit in a manner very similar to how a supersonic {\em gas} jet would.  That is, somewhere between the black hole jet region and the lobe jet region, a jet like that in Cygnus A (\textbf{Figure~\ref{fig:cyga}}) must have lost much of its electromagnetic energy.  Most of the energy dissipated in FR-II lobes, therefore, appears to be kinetic -- from a still highly-relativistic flow.  

Both FR-I and II jets dissipate much of their energy in the lobes. However, the former FR-Is appear to lose much of their kinetic energy in or near the black hole jet phase, but keep much of their electromagnetic energy well beyond. The latter FR-IIs behave oppositely, losing much of their electromagnetic energy in or near the black hole jet region, but keeping much of their relativistic kinetic energy out to the lobes.  

\subsection{Numerical Simulations of Jets}\label{ssec:rmhdsim}
Much of our current understanding of the formation of relativistic jet derives from 3D GRMHD
\begin{marginnote}[2pt]\entry{GRMHD}{General Relativistic Magneto-Hydrodynamics}\end{marginnote}
simulations \citep{meier12,nakamura18} though much is also being learned from special relativistic and lower dimensionality simulations. Many simulations also include radiative transfer and dissipative processes \citep{18,mckinney14}.

\subsubsection{Short History}
Some historical simulations are still worthy of note.  \citet{norman82} performed a 2DHD simulation of a supersonic jet propagating through a uniform density medium.  The detailed results confirmed the \citet{blandford74} jet hot spot picture of a Mach disk, contact discontinuity, and bow shock forming the jet head, with hot post-shocked jet material surrounding the advancing jet.  This picture readily explains the lobe structure of Cygnus A and other FR-II sources.

\citet{lind89} and \citet{clarke86} performed similar simulations, but with a 2DMHD code and a strong helical magnetic field encircling the jet.  Again, a Mach disk/bow shock structure formed, but there were two distinctive differences.  The shock appeared to regularly eject supersonic components/shocks downstream, and the shocked jet material formed a downstream nose cone or ``plug'' ahead of the original jet.  Such behavior was not seen in hot spots of FR-II sources many kiloparsecs from their SMBHs, but it appears to be similar to time-dependent evolution of parsec-scale VLBI sources -- in a possible recollimation shock near $R_{\rm inf}$.  \citet{mignone10} performed 3DRMHD simulations of a similar situation and found that the helically-magnetized jet develops a strong kink instability, which inhibits the forming of a nose cone.  However, there are several ways in which nature can avoid or mitigate the kink instability in Poynting jets \citep{nakamura04}: the jet is propagating through a decreasing-pressure atmosphere or the jet plasma is spinning faster than its internal Alfv\'{e}n speed. Indeed, the M87 black hole jet is clearly Poynting-dominated, accelerating and collimating over $\sim 300$ pc, and yet does not kink until well beyond HST-1. 

\subsubsection{Jet Launching}
\paragraph{From Sub-Eddington Accretion Disks}\label{par:subed}
Most simulations of jet launching involve radiatively inefficient (very sub-Eddington) accretion flows (RIAFs) around rotating black holes -- the type that one would expect in the engines of FR-I and weaker sources.  From the {\sl tour de force} 2D and 3D GRMHD simulations \citep{gammie03,devilliers03,mckinney06}, it is clear that accreting, rotating black holes can produce jets and that the jet power is an increasing function of the SMBH spin.

\paragraph{From Super-Eddington Accretion Disks}\label{par:suped}Rather rare sources that are thought to be accreting above the Eddington limit, {\textit e.g.} broad absorption line (BAL) quasars, are known to be less radio loud than the main QSR population -- but not radio silent.  It would seem, then, that even with a large, geometrically thick, radiation-pressure dominated disk, such objects still can produce jets.  These ideas have been investigated by \citet{sadowski14,sadowski15,mckinney14,mckinney15} using a 3D GRMHD code that includes radiation magnetohydrodynamics of disks with $\dot{m} \sim 20$ and $50$. They find that the radiation and jet emerge from a geometrically beamed, bi-polar region, with super-Eddington isotropic luminosities.  (That is, the total radiation and jet luminosity is of order Eddington, but much of it is beamed in a narrow polar funnel.)  As expected, the super-Eddington wind carries away a significant amount of jet energy, but nevertheless a jet is produced.  In addition, the heart of the jet engine can be seen down the funnel by an observer if the viewing angle is small enough.  

\subsection{Jet Confinement}\label{ssec:jetcon}
A longstanding question about AGN jets is how are they confined and collimated? This question likely has different answers at different radii. If we accept that the jet originates from close to the black hole, then it has been most commonly assumed that the jet is initially confined by the walls of a funnel formed by an ion-supported torus. However, this may not be present and, in any case cannot extend to large distance where we still see jets being collimated. A gas dynamical disk wind may confine the black hole jet close to the hole \citep{globus16}.  Alternatively, if the wind is hydromagnetic, then the jet may carry an axial current $I$ which supports a toroidal magnetic field $B_\phi\sim0.3(I/1\,{\rm EA})(r/1\,{\rm pc})^{-1}\,{\rm G}$, for $r\ge r_{\rm jet}$ \citep{levinson17}. Ultimately, at large enough cylindrical radius, $r_{\rm out}$ there will be a return current and the outward magnetic stress must be balanced by gas pressure \citep{begelman84}. However this pressure will be $\sim(r_{\rm out}/r_{\rm jet})^{-2}$ smaller than the pressure in the jet. As discussed, the galaxy jet is probably directly confined by gas pressure, and in the lobes in the case of FR-II sources, by the back flow of jet plasma that has passed through the hot spot shock.    

\subsection{Jet Propagation and Shock Behavior}\label{ssec:jprop}
As we discussed above, multidimensional simulations with a strong helical magnetic field are particularly applicable to the structure and evolution of parsec-scale radio sources, especially near the jet collimation break (\textbf{Figure~\ref{fig:m87rR}}). As further evidence, 1.5DRMHD jet simulations, also with a strong helical magnetic field, have been applied to the time-dependent structure of HST-1 in M87 \citep{nakamura10}. In contrast to hydrodynamic simulations of jet flow, however, if the jet speed is super-magnetosonic in both the ambient medium and internally in the jet flow itself, the 1.5RMHD results show that {\em two} bow shocks and {\em two} Mach disks form in the flow.  The leading bow shock is a fast-mode magnetosonic shock and the one behind it is a slow-mode magnetosonic shock.  Conversely, the leading Mach disk, somewhat behind the forward-slow bow shock, is a reverse-slow shock. And, finally, a reverse-fast Mach disk trails the reverse-slow Mach disk.  This quad-shock system is seen in HST-1 outbursts and can be fit with just a few parameters to a super-magnetosonic jet impacting a slower one from behind \citep{nakamura10}.  

Furthermore, the shock angular momentum conservation condition induces plasma rotation between the two bow shocks and then also between the two Mach disks, with the two inter-shock regions rotating in opposite directions.  Such a behavior may explain oppositely-directed polarization rotations in blazars like OJ287, which has a viewing angle of only 2 degrees \citep{cohen18}.  

\section{EMISSION MODELS}\label{sec:emmod}
\begin{extract}
``The observations by Baade (1956) of the polarization of the jet in M87 are convincing proof of the hypothesis that this optical radiation is synchrotron radiation emitted by relativistic electrons and positrons moving in magnetic fields\dots'', ``The jet will also be a source of high- energy gamma radiation.'' -- From \citet{burbidge56}.
\end{extract}

\subsection{Particle Acceleration}\label{ssec:paccn}
\subsubsection{General Principles}\label{sssec:paccnP}Jets, and the double radio sources that they supply, are observed throughout much of the entire seventy octave electromagnetic spectrum and, perhaps, beyond. Even if a jet is well-modeled at the level of continuum mechanics, we must still explain where and how the emitting relativistic particles are accelerated and how they are transported before they cool. The range of electron energies for which we have good evidence range from $\lesssim100\,{\rm MeV}$ to explain lower brightness, and self-absorbed radio cores to $\gtrsim100\,{\rm TeV}$ to explain X-ray synchrotron  and $\gamma$-ray emission. Much higher energies are invoked in some models (\textbf{Section~\ref{ssec:crnu}, \ref{par:pgpi}}). The case for some distributed acceleration along the jet is very strong as adiabatic and radiative cooling will ensure that the surface brightness falls much faster with $R$ than observed.

A charged particle gains energy at a rate $eEc$ when exposed to a electric field of strength $E$.  If the field varies slowly, the acceleration is called electrostatic. This is likely to occur in black hole magnetospheres where charged particles must be created continuously and electrostatic gaps are thought to be involved, though they are likely to be energetically insignificant as the potential difference needed is orders of magnitude smaller than the total available \citep{levinson17a,chen18}. Similar structures have been invoked in the black hole jet and may be mandated as the density of charge carriers continues to be inadequate to deliver the current density and space charged required by Maxwell's equations. However, entrained gas will soon make this a non-issue and electrostatic acceleration is unlikely to explain most of the observed jet emission. When  there are many more protons than positrons, then they should take up nearly as much energy as the electrons while they can be accelerated to much higher energy as their radiative losses are less.

\subsubsection{Diffusive Shock Acceleration}\label{sssec:dsa}Acceleration at high Mach number shocks has often been invoked as the primary particle acceleration mechanism in jets. 
This is observed to be very efficient in supernova remnants and at planetary bow shocks. In its simplest version, a strong planar shock with compression ratio $r$ transmits a downstream momentum space power-law distribution function in momentum with slope $3r/(r-1)$ so that a strong shock with $r\sim4$ will lead to a synchrotron or Compton power law with $\alpha\sim-0.5$ as is often observed \citep{drury83}. Weaker shocks can account for steeper spectrum sources. 

The shock should be non-relativistic and the plasma $\beta>>1$
\begin{marginnote}[2pt]\entry{$\beta$}{The ratio of a pressure or a partial pressure to the magnetic pressure.}\end{marginnote}
ahead of the shock and must be corrected to allow for the dynamical effect of the relativistic particles on the shock structure which can decelerate the upstream flow and change the effective heat ratio. Low $\beta$ shocks are not compressive and relativistic shock acceleration probably converts most of the downstream energy into thermal (relativistic) particles. The shocks can accelerate to high energy because magnetic field is also created at the shock front by the accelerating particles. The maximum energy accelerated should be determined by the shock width or radius of curvature. A further complication is that the very highest energy particles can escape upstream from a curved shock front, although most accelerated particles will be transmitted downstream and can lose energy during subsequent expansion and radiation. Despite these limitations, shocks remain a good candidate for particle acceleration in galaxy jets, the hot spots and the radio lobes. In particular, recollimation shocks near $R_{\rm inf}$ may be efficient accelerators \citep[e.g.][]{marscher14}.  

\subsubsection{Relativistic Reconnection}\label{sssec:magrec}
Axisymmetric relativistic jets confined by toroidal magnetic field are prone to non-axisymmetric kink instabilities that can force oppositely directly field lines together so that they reconnect by exchanging partners \citep[\textit{e.g.}][]{begelman98,duran17}. This will occur in many comparatively small regions where ohmic dissipation balances flux-freezing. Non-relativistic reconnection is rather inefficient with most of the dissipated energy going into heating the thermal plasma, although surprisingly large energies may be accelerated. Relativistic reconnection has been investigated extensively using PIC
\begin{marginnote}[2pt]\entry{PIC}{Particle In Cell}\end{marginnote}
simulations and can be very efficient in converting magnetic energy, measured in a comoving frame into ultrarelativistic particles \citep{sironi14,werner18}. 

Simulations of current sheets, which form in magnetized jet flows, show them breaking up into a series of islands separated by X-points \citep{beloborov17}. Channels can be formed by the electric vector that can be locally larger in magnitude than the magnetic field, electrons and positrons are accelerated along the field and then radiate synchrotron radiation at the ends of the channel where they encounter a strong transverse magnetic field. Other acceleration modes have been associated with relativistic reconnection, including scattering within and between islands and small pitch angle acceleration along a guide field lying in the main current sheet. Reconnection should be especially prominent in the black hole jet boundary layer. Synchrotron radiation from these regions should exhibit high, perpendicular polarization.

Shock and reconnection acceleration are associated with surfaces --- current sheets. Continuous emission across and along jets and lobes should be observed so long as the electrons do not cool on the shorter of the flow and propagation timescales. However dynamical features where the field is expected to be strong and the particle density high, should also be highlighted. In particular the knots in sources like M87 and the hot spots in FR-II lobes are plausibly shocks, while reconnection should be endemic in boundary layers. Shocks and reconnection should be distinguishable through X-ray polarimetry \citep{tavecchio18}

\subsubsection{Stochastic Acceleration}\label{sssec:magcon}
Supersonic gas dynamical jets are famously noisy, which means that they radiate sound waves. Similar MHD wave emission is expected in AGN jets. These waves are likely to accelerate particles through second order, stochastic processes and mediate a form of local viscosity in the flow \citep{kulsrud04,stawarz08}. (Note that a high level of turbulence is likely to inhibit particle transport especially of radio-emitting electrons with small gyro radii. It can, however, promote reconnection.) Wave acceleration is more likely to be volumetric, contributing to emission within the core of a jet, especially if jets develop large-scale turbulent eddies. It should also be especially effective in boundary layers though the resulting synchrotron radiation is likely to be less strongly polarized than with relativistic reconnection.

\subsubsection{Magnetoluminescence}\label{sssec:maglum}
While these processes may suffice to account for slowly varying emission from jets, they may not be fast enough to explain the most dramatic $\gamma$-ray variation seen in jets and other sources like the Crab Nebula. One possible candidate mechanism is suggested by observations of solar prominences and, perhaps, the Galactic center. The magnetic field in a low $\beta$ jet may organize itself into current-carrying ropes that become tangled as the field lines are pulled and twisted by the large scale jet flow. Discrete ropes can untangle, without topological change at speeds $\sim c$, creating a giant volumetric inductive electric field. The reduction in the magnetic energy should appear as high energy electrons and ions created on the light crossing time of the reconnecting region \citep{blandford17}. 

\subsection{Radiative Processes}\label{ssec:radn}
The familiar view of jet emission is that their spectra commonly appear to have a Bactrian form with two humps (\textbf{Figure~\ref{fig:blseq}}) -- the lower peak being associated with synchrotron emission, the higher one with Compton scattering of synchrotron photons in low power sources and external disk and broad emission line cloud photons in high power sources (\textbf{Section~\ref{ssec:gam}}). The simplest, one zone, version of this model is that both humps originate from a single site, plausibly associated with $R_{\rm inf}$, where a re-collimation shock may form (\textbf{Sidebar Radiative Transfer}). A more complex model brings in variation along the jet. While the Compton--synchrotron model may still capture the main observed features of AGN jets, there are some challenges

\begin{textbox}[ht]
\section{RADIATIVE TRANSFER}
\subsection{Synchrotron Radiation}
A relativistic electron (or positron) with energy $\gamma m_e c^2$  spiraling along a uniform magnetostatic field $B$ (measured in G in a reference frame with no electric field) emits linearly polarized ($\sim70$~\%) synchrotron radiation with characteristic frequency $\nu\sim\gamma^2B\,{\rm MHz}$. The electron cooling time is $t_{\rm syn}\sim5\times10^8B^{-2}\gamma^{-1}\ {\rm s}$ \citep{rybicki79,longair11}. The degree of circular polarization is $\sim3/\gamma$ for electrons alone. Absorption can be estimated by noting that the brightness temperature satisfies a thermodynamic limit $3kT_B=3Ic^2/2\nu^2\lesssim\gamma m_ec^2$. When transforming from the source frame to the observer frame, $T_B$ transforms in the same manner as $\nu$. Also relevant is the possibility of absorption of radio emission by cool plasma along the line of sight, for example in an accretion disk. The absorption coefficient for radio free-free emission can be written as $\mu_{\rm ff}\sim10^{-25}(<n_e^2>/{\rm cm}^{-6})(T_e/10^4\ {\rm K})^{-1.35}(\nu/1\ {\rm GHz})^{-2.1}{\rm cm}^{-1}$.

\subsection{Faraday Rotation}
Linear polarized radio waves that propagate through a magnetized, electron-ion plasma can be decomposed into circular polarized eigenmodes which propagate with slightly different phase velocities so that the plane of polarization appears to rotate after they are combined at the end of the propagation by an amount $RM\lambda^2$, where the rotation measure is $RM=8.1\times10^5\int(ds/1\,{\rm pc})(n_{\rm e}/1 \,{\rm cm}^{-3})(B_{||}/1\,\mu{\rm G})\, {\rm radians\, m}^{-2}$ with $n_e$ being the electron density, and $B_{||}$ the parallel component of magnetic field. The wavelength, $\lambda$--dependence of the rotation allows the product of the field and the density to be estimated around jets. 
\subsection{Inverse Compton Scattering}
A relativistic electron (or positron) in a jet can collide with an ambient soft photon according to the Thomson cross section, $\sigma_{\rm T}\sim7\times10^{-25}\ {\rm cm}^2$. The photon may be emitted by the jet itself, by stars, by the accretion disk or by gas clouds. The photon frequency/energy is Doppler-boosted once from the frame of the soft photon to the electron rest frame and then a second time from the rest frame back to the original frame by an average factor $\sim\gamma^2$ with the numerical coefficient dependent upon the angular distribution of soft photons. This suffices to compute the emissivity with the caveat that if the scattered photon energy approaches that of the electron, Compton recoil is important and the emissivity is diminished. Below this ``Klein-Nishina limit'', if we define an effective magnetic field strength which would have an energy density equal to that of radiation field then the electron cooling time is similar to that for synchrotron radiation. The requirement that the $\gamma$-ray flux from the radio source in the jet not exceed what is observed  limits the synchrotron radio brightness temperature to $\sim10^{12}\ {\rm K}$, ignoring Doppler-boosting.
\subsection{Pair Production and Annihilation}
When a $\gamma$-ray with energy $h\nu_\gamma$ encounters a soft photon head on with energy $>m_ec^4/h\nu_\gamma$, there is enough energy to create an electron-positron pair. The cross section for this process is $\sim0.2\sigma_T$.In practice this is provides a lower bound on the radius where $\gamma$ rays are produced in an AGN jet. The inverse process is annihilation of electrons and positrons which proceeds with a cross section $\sim\sigma_T(c/v_e)$, where $v_e$ is the electron thermal speed.
\end{textbox}

\subsubsection{Some Current Questions}\label{sssec:quest}
\paragraph{Is the radio emission synchrotron radiation?}\label{par:qsyn} 
The radio brightness temperatures should satisfy $T_{\rm B}\lesssim10^{12}{\cal D}/(1+z)\,{\rm K}$ in order to avoid catastrophic production of Compton $\gamma$-rays \citep{begelman84}. This is a strong constraint when the source is resolved and the brightness temperature is directly measured and there may be instances when it is not satisfied. It is less secure when the linear size is estimated by variability which may be attributable to relativistic kinematics or scintillation.
\paragraph{Is the $\gamma$-ray emission synchrotron radiation?}\label{par:gsyn}
If we suppose that the electric field strength does not exceed the magnetic field strength, in cgs units, then electron synchrotron radiation photons must have wavelengths more than the classical electron radius, or energies $\lesssim70\,{\rm MeV}$ \citep{landau75}. However, if electrons are created hadronically at very high energy, they can simply cool, emitting more energetic photons. 
\paragraph{Can jets transition rapidly from electromagnetic to particle dominance?}\label{par:qem} 
In many sources, the flux from the Compton peak significantly exceeds that from the synchrotron peak implying that $\beta_{\rm rad}>>1$ and $\beta_{\rm e}>>1$. By contrast, we have emphasized a scenario in which jets are powered electromagnetically by a spinning black hole where the opposite inequality should be satisfied. 
\paragraph{Do $\gamma$-rays originate from jet radii $\lesssim1000 \,r_{\rm g}$ or are there bulk Lorentz factors $\gtrsim100$ in the flow?}\label{par:qgr} 
Variation in some $\gamma$-ray jets with timescales $\lesssim\tau_{\rm g}$ (\textbf{Section~\ref{ssec:gam}}) suggests that much larger Lorentz factors are present than indicated by observations of superluminal motion and the emitting volume is very compact, while still being large enough to account for the observed flare fluence. 
\paragraph{Is there something seriously wrong with our basic model of accretion in AGN?}\label{par:qmod}
 There is a firm lower bound to the radius from which $\gamma$-rays of energy $E_\gamma$ can escape called the gammasphere. This is the surface where the optical depth to pair production on photons with energy $\gtrsim m_e^2c^4/E_\gamma$ is unity, where the cross section $\sim0.2\sigma_{\rm T}$.  Geometrical corrections can soften this constraint somewhat and, in the case of BLL, we can argue that the accretion disks are dark. However, the radiation field is well-measured in a quasar, like 3C279, which displays minute scale GeV variation (Figure~\ref{fig:3c279}). 

\subsubsection{Alternative Emission Models}\label{sssec:altem}
In view of these problems, several alternative emission models have been entertained.
\begin{marginnote}[2pt]\entry{SED}{Spectral Energy Distribution}\end{marginnote}  
\paragraph{Proton Synchrotron Radiation}\label{par:psyn} This has been associated with compact radio sources, X-ray jets and hot spots mainly because the lifetime of a particle with mass $m$, emitting at a given frequency in a given field $\propto m^{5/2}$ \citep{petropoulou12}. Although short-lived in their rest frames, ultrarelativistic pions and muons produced by hadronic processes can also contribute to $\gamma$-ray synchrotron emission.  
\paragraph{Curvature Radiation} The challenges of high brightness electron synchrotron sources can be obviated by invoking coherent emission processes. In particular, curvature radiation in which bunches of electrons stream long curving magnetic field produced brightness temperatures limited by the total energies of the bunches --- a mechanism that has been invoked in pulsars.  A serious constraint that these models should satisfy is that high brightness emission not be degraded by nonlinear transmission effects including induced Compton scattering and stimulated Raman scattering \citep{levinson95}. Curvature radiation has also been invoked to explain $\gamma$-ray emission \citep{bednarek97}.  
\paragraph{Cyclotron Masers}\label{par:cycm} Alternatively the brightness temperature limitation can be broken with relativistic electrons if there is a population inversion and the emissivity declines with increasing particle energy. This can happen with cyclotron emission by mildly relativistic electrons due to the dependence of the cyclotron frequency on the relativistic mass. This is observed in Jovian decametric emission \citep{begelman05}.
\paragraph{Plasma Waves}\label{par:plwa} Shocks and magnetic discontinuities can produce large current densities which may radiate large amplitude plasma modes which may, in turn, mode-convert into electromagnetic radio waves as seen in solar radio bursts.
\paragraph{Proton-pion Production by Energetic Protons}\label{par:pppi} Another widely discussed possibility for the $\gamma$-ray emission is that it is hadronic \citep{muecke03,reimer12}. The basic idea is that most of the energy dissipated in the jet goes into protons (not pairs) and that these are accelerated relatively rapidly to PeV or even EeV energies where they can undergo a variety of interactions and create showers like cosmic rays hitting our atmosphere. The first processes considered were p-p collisions which could happen if, for example, a jet moved to encounter a dense molecular cloud. 
\paragraph{Photo-pion Production by Energetic Protons}\label{par:pgpi} Pions may also be produced by collisions with photons \citep{muecke03}. To give one example, consider a FSRQ with spectrum peaking at $\sim30\,{\rm meV}$ or roughly $30\,\mu{\rm m}$. Suppose that the jet has a Lorentz factor  $\Gamma\sim10$ so that the background photons have energy $\sim10\,{\rm meV}$. Now suppose that protons can be accelerated energies $\sim500\,{\rm PeV}$ in the jet frame. The photons will have energies $\sim150\,{\rm MeV}$ in the proton rest frame, just above the threshold for pion production which will proceed with a cross section $\sim10^{-4}\sigma_{\rm T}$. Roughly $m_{\rm p}/m_\pi\sim10$ pions must be made to extract all the proton energy. Charged pions will decay into muons, neutrinos, electrons and positrons which can emit synchrotron and Compton radiation. Typical electron energies are $\sim100\,{\rm PeV}$ in the AGN frame. The neutrinos are relevant to their possible detection (Section~\ref{ssec:crnu}). Neutral pions will decay directly into $\sim1\,{\rm EeV}\ \gamma$-rays which will quickly cascade into pairs and lower energy $\gamma$-rays until the latter can escape. The neutron will escape the source before it decays back to a proton.  Overall, the process is quite inefficient as the cross section is low and, typically, no more than a few percent of the initial proton energy will emerge as $\gamma$ rays for each pion created. Also, this could raise the integrated density of SMBH to unreasonable levels \citep{yu02}. 
\paragraph{Photo-pair Production in a Shielded Jet}\label{par:pgpair} In order for $\sim1\,{\rm GeV}$ photons to escape from close to the hole in a FSRQ and vary rapidly, it is necessary to exclude UV and soft X-ray photons. One way to do this is if there is a disk wind at small radius driven by radiation and magnetic field that forms a sheath around the inner jet that can absorb essentially all of the external photons shortward of the Lyman continuum \citep{meyer18}. The mass flow needed is modest but the gas would have to cool fast enough to have a large enough neutral fraction \citep{konigl94}. Under these circumstances, the hardest photons would have energies $\sim100\,{\rm eV}$ in the jet frame. The cross-section for photo-pair production rises from zero at threshold, when the photon energy is $\sim1\,{\rm MeV}$ in the proton rest frame to $\sim0.01\sigma_{\rm T}$ when the energy is $\sim100\,{\rm MeV}$ before pion production sets in. An accelerating proton will therefore be subject to increasing radiative drag as its comoving frame energy approaches $\sim1\,{\rm PeV}$ and $\sim50\,{\rm TeV}$ pairs will be produced.  If the jet is highly magnetized at this radius, as we are otherwise assuming, the pairs will cool rapidly by synchrotron emission of photons with energy $\sim10\,{\rm GeV}$ in the AGN frame. These are just able to escape. As the photon density and cross section are both larger than with pion production, it is possible to convert half the electromagnetic energy dissipated into GeV $\gamma$-rays from the inner jet with very high efficiency. 

\subsection{Jet Emission Profile}\label{ssec:empro}
We are a long way from being able to model the observed intensity distribution of AGN jets and, instead use it, in combination with simulations, to try to infer the physical conditions and the acceleration and emission processes. Despite this, there are some key effects that have either been seen or should now be be accessible to observation.

\subsubsection{Radio Core-shift}\label{sssec:cshift}
The unresolved radio core has generally been associated with self-absorbed synchrotron radiation from the approaching jet. We are, in effect looking at a radio photosphere formed by the jet walls and a surface across the jet. The size of the unresolved photosphere should scale roughly as the wavelength and the size that can be resolved interferometrically with an inter-continental baseline. This expectation is consistent with multi-frequency VLBI observations. However, we expect to see the unresolved centroid of the photosphere shift towards the position of the black hole with increasing $\nu$ \citep{blandford79,konigl81}. This effect has been seen \citep{plavin18}.

\subsubsection{Radio-$\gamma$-ray Cross Correlation}\label{sssec:rgcore} 
The statistical evidence that the $\gamma$-rays vary before the radio emission suggests that the variable $\gamma$-ray emission site lies within the radio core (\textbf{Section~\ref{ssec:var}}) discrimating between models.

\subsubsection{$\gamma$-sphere}\label{sssec:gsph}  
Likewise, if the $\gamma$-ray variation is dominated by activity at the $\gamma$-sphere, with radius increasing with photon energy, then we would expect low energy variation to precede high energy changes. Alternatively, if it is due to hadronic showers in an optically thin region, then we might expect the opposite ordering. However, there is not yet strong evidence either way.

\subsubsection{Knots in M87}\label{sssec:knots} 
Important clues concerning general jet dynamics and radiation properties have have been drawn from a few well-resolved X-ray and optical jets (Sec.~\ref{ssec:oiru}). The six M87 knots have been interpreted as strong, particle-accelerating shocks which would indicate that $\beta\gtrsim1$ \citep{biretta99}. They have also been interpreted as places where the jet velocity turns towards us \citep{bicknell96}. Given the successful observing campaign involving  HST-1 it may be possible to extend this to the innermost knots.

\subsubsection{Rapid variability}\label{sssec:rapvar}
Variability from optically thick radio emission through TeV $\gamma$-rays provides important constraints on physical conditions within jets as the associated timescales are sometimes $\lesssim r_{\rm g}/c$ which has motivated discussion of extra Lorentz factors and spatial focusing of jets. Symmetric and non-symmetric variations, lags and the orphan flares have been identified \citep{potter15}.

\subsubsection{Lessons from Double Radio Sources}\label{sssec:double} 
There is also much to be learned from observations of extended radio sources that may be scalable to smaller scale jets and Compact Symmetric Objects (Section~\ref{sssec:lobes}). Comparison of X-ray and radio images of Cygnus A are essentially hydrodynamical with $\beta>>1$ and are able to account for the jet-terminating hot spots, which dance about like a dentist's drill \citep{scheuer82}, as well as the back flow and a large cocoon enveloping the whole galaxy in  pressure equilibrium with the surrounding gas and permeated by weak shock fronts. However the large linear polarizations reported in many lobes suggest that magnetic stress is important and $\beta\sim1$.

\section{AGN Jets and the Universe}\label{sec:uni}
\begin{extract}
``The value of $N / N_m$ found for our sample of thirty-three 3CR sources is 0.75. This in direct conflict with the value of 0.50 required by a steady state [universe] in which no evolutionary effects are admitted.'' -- From \citet{schmidt68}; discovery of radio source population evolution ruled out the steady state cosmological model in favor of the big bang.
\end{extract}

\subsection{Radio-Loud / Radio-Quiet Dichotomy}\label{ssec:rlrq}
The classification of AGN, like the classification of stars, has evolved with the compilation of multi-wavelength data and partial success in relating the taxonomy to physical characteristics of the sources \citep[\textit{e.g. }][]{meier12}. Essentially all bright galaxies contain SMBH which may become active. Now, roughly ten percent  of optically-selected quasars are RLQ \citep{ivezic02,kellermann16}. Jets are not always observed in these sources but are presumed to be present. The remaining  roughly ninety percent are RQQ, though not silent \citep{barvainis05}. This, the simplest classification of active galaxies, is summarized in \textbf{Table~\ref{tab:ROQ}}. 

From the perspective of optical spectroscopy, AGN are characterized by their line emission which is commonly split into broad and narrow components \citep{hine79,jackson99,osterbrock05}.   The absence of broad lines from Sy 2/NLRG is attributed to obscuration of a broad line region by dusty gas associated with a thick or heavily warped outer disk, or torus, an outflowing wind or inflowing gas. Therefore, these objects are those for which the observer inclination, $\theta$, is large with respect to the black hole spin, presumed here to be aligned with the disk angular momentum and the jet, when present \citep{urry95,meier12}.  Weak, polarized, back-scattered broad lines are observed from Sy 2/NLRG, which supports this explanation \citep{antonucci85}. Optically, RLQ, RQQ are intrinsically similar with similar masses and accretion rates.

One possible reason for a quasar forming a powerful jet and becoming radio loud is that the SMBH is spinning rapidly. Observations of black hole X-ray binaries support the view that spin is necessary for jet production \citep{narayan12}.  However, many Seyferts with large measured spin are radio quiet \citep{reynolds14} and, using the \citet{soltan82} argument, \citet{yu02} and \citet{elvis02} have argued that most RQQ spin rapidly. It appears that black hole spin is also insufficient for jet production. This is referred to as the Spin Paradox. It implies that a second, independent factor must be invoked to explain why a fraction $\sim10^{-4}$ of local AGN rising to $\sim0.1$ of high redshift quasars should produce jets and become radio-loud largely independent of how the optical emission is produced. 

One explanation (Section~\ref{sssec:accstate}), which underpins most numerical simulations, is that the gas supply rate is either too slow or too fast to allow the accreting gas to cool and form a large torus and funnel to collimate the jet. A longstanding concern with this explanation is that it requires the arrangement to persist for $\gtrsim10\,{\rm Myr}$, the lifetime of a radio source. Given the observed rapid variability in AGN, it seems hard to believe that the  accretion rate is that stable. A second worry is that we now observe jets being collimated at altitudes, $r_{\rm g}<<r<<R_{\rm inf}$, that are too large for a thick accretion disk to be important and too small to be associated with the interstellar medium.

\subsubsection{Disk Magnetization}\label{sssec:diskmag}
\paragraph{Near Magnetization}\label{par:nearmag}The existence of the MRI, ensures that disks are strongly magnetized. However, it does not guarantee that there is a permanent, large scale vertical field threading the disk and the SMBH in RL and not in RQ AGN.  A flux large enough to produce a MAD
\begin{marginnote}[2pt]\entry{MAD}{Magnetically Arrested Disk, where the central magnetic flux holds back inflowing gas}\end{marginnote} 
or a MCAF
\begin{marginnote}[2pt]\entry{MCAF}{Magnetically Choked Accretion Flow, where inflowing gas leads to an enhancement of hole magnetization}\end{marginnote} 
will hold up the flow outside the $r_{\rm ISCO}$, generate more powerful jets and may even be strong enough to suppress the MRI for a while \citep{narayan03,mckinney12}.  One variation, reviewed by \citet{meier12}, is that the disk is counter-rotating with respect to the spin in RL AGN, allowing more flux to accumulate within the ISCO. However, this conjecture was not supported by simulation \citep{tchekhovskoy11,tchekhovskoy12}.

Stellar black holes like GRS 1915+105 \citep{mirabel99} can provide further insight. This source produces powerful, intermittent jets when the accretion rate is near Eddington and when the source passes through the ``jet line'' in the X-ray hardness-intensity diagram, which it does every 20~min with a duty cycle $\sim0.06$ \citep{fender04}. This is attributed to a disk instability causing a hot, ADAF-like structure which persists for an inflow timescale and is capable of creating a large field through dynamo action \citep{meier12}. Unfortunately, there is no evidence of this cycling on timescales which should increase in proportion to mass to thousands of years in AGN jets. Furthermore the fact that being radio-loud or -quiet appears to be a long term choice makes it hard to associate with instabilities that grow on a dynamical timescale close to the black hole.

A rather different mechanism that may operate in the presence of radiation pressure, and which addresses this problem, is that a physical dynamo continually regenerates field of fixed polarity  --- the sign of $\vec\Omega\cdot\vec B$ --- near the inner accretion disk \citep{contopoulos18}. This could be due to radiation pressure acting on electrons combining with the MRI, which expresses no such preference. This should be testable. 

\paragraph{Distant Magnetization}\label{par:distmag}A possible clue is that radio sources are mostly associated with elliptical and not spiral galaxies. This suggests that it is the environment at radii $r\gtrsim R_{\rm inf}$ that is responsible for the dichotomy. The large scale gas inflow in a spiral is likely to be mostly  equatorial and it may be hard to build up and trap magnetic flux within $R_{\rm inf}$ as the field lines become radial on the surface of the galaxy disk and may quickly reconnect and escape. 

By contrast, in an elliptical galaxy, the inflow derives from the circum-galactic medium and should be quasi-spherical. Magnetic flux  can  accumulate and be effectively trapped within a large funnel with radius $\gtrsim R_{\rm inf}$ formed by hot gas settling onto the nucleus at a rate $\dot M$, and extending to high latitude. The magnetic stress at $R_{\rm inf}$  would balance the ram pressure of the infalling gas. The accretion disk would wind up and the field and making the very simplest assumptions, the electromagnetic power of the spinning black hole is $\sim\dot M\sigma cj^2$. A similar power  would be associated with a hydromagnetic wind leaving the inner accretion disk. The actual accretion rate onto the black hole could be much smaller than the infall rate.

\subsection{Fanaroff--Riley Classification of Extended Radio Surces}\label{ssec:fr}
\subsubsection{Compact and Extended Radio Souces}\label{sssec:compext}
We have developed the view that compact and extended radio sources are intrinsically similar but in the former case a relativistic jet is pointed towards us and in the latter, it is not. As predicted, compact radio sources, more generally, blazars, have faint extended halos, their unbeamed lobes. Likewise, extended radio sources have faint cores identified with the nuclei of their host galaxies.  This bold unification implies that both types of source have similar distributions of intrinsic jet and environmental properties.

\subsubsection{Physical Origin of the FR Classification}\label{sssec:frtrans}
The Fanaroff-Riley class of the extended sources (Section~\ref{ssec:hist}, \textbf{Table~\ref{tab:RLAGN}}) remains an impressively striking feature of the observations. Physically, what seems to be happening is that jets with power $\gtrsim5\times10^{45}\,{\rm erg\,s}^{-1}$,\citep{potter15}, are able to escape the galactic nucleus with little deceleration by the surrounding and confining medium or a magnetic sheath, a slower disk wind or a settled interstellar medium. The opposite is true for lower power jets and once they start to decelerate, they are quickly converted into subsonic plumes that fade with radius and inflate buoyant bubbles (\textbf{Figure~\ref{fig:m87mon}}). We know that FR-I jets are initially relativistic and that the observed characteristics that determine if a blazar is classified as a BLL or a FSRQ emanate from radii $\lesssim R_{\rm inf}$ \citep{kharb15}. We also observe that when the transition occurs, it does so far inside the galaxy's core radius \citep[cf.][]{bicknell95}. The existence of hybrid morphology radio galaxies --- an FR-I jet and an FR-II jet --- also supports FR-I jets being formed near $R_{\rm inf}$. 

From a fluid point of view, $R_{\rm inf}$ is a very natural location to change the nature of a flow as there is an abrupt change in the gradient of the gravitational potential and, therefore, in the external pressure gradient \citep{levinson17}. A recollimation shock may form and disrupt a low power jet \citep{asada12}. Alternatively, a highly magnetized jet may be subject to a disruptive kink or helical instability \citep{tchekhovskoy16}.

The major change in a FR-II jet is that it appears to be strongly magnetized when formed while the best studied hot spots where it terminates are best modeled by a weakly magnetized upstream flow that accelerates electrons up to $\sim1\,{\rm TeV}$ and amplifies the field up to $\sim100\,\mu{\rm G}$ \citep{werner12,araudo18}. The dynamical decline of the jet magnetic field could be abrupt if recollimation induces turbulence which facilitates magnetic reconnection and dissipation \citep{meier13}.   

\subsection{Blazar Sequence}\label{ssec:blseq}
In the optical, BLL show, at most, weak lines while FSRQ have strong, broad lines. This is usually attributed to a change in the accretion mode from thick, radiatively inefficient inner disks to thin radiatively efficient disks radiating most of their power in the UV although there is a possibility that radiative inefficiency is due to dominant, external magnetic torques extracting most of the angular momentum (\textbf{Sidebars Black Hole Accretion, Electromagnetic Effects}). 

When we turn to the beamed jets, the BLL/FSRQ are generally unified with beamed FR-I/II sources (\textbf{Section~\ref{ssec:hist}, Tables~\ref{tab:RLAGN}, \ref{tab:BLSEQ}}). Curiously, BLL, subdivided into four subclasses and FSRQ differ significantly in their radio, optical and high energy properties and can be arranged in a sequence (see \citet{fossati98}, \textbf{Figure~\ref{fig:blseq}}). In the radio, the BLL magnetic field is perpendicular to the projected jet axis, even when the jet bends; FSRQ magnetic field is either aligned or random \citep{gabuzda14}. This is argued to imply that BLL jets contain a strong, helical, magnetic field, while the emission from FSRQ are dominated by either turbulent or laminar shear flows \citep{hughes89}. 

 We have contrasted leptonic and hadronic models of the $\gamma$-ray peak in FSRQ (\textbf{Sections~\ref{ssec:gam}, \ref{par:pgpi}, \ref{par:pgpair}}) and argued that the latter may be able to account for rapid GeV variability if a disk outflow can shield the jet efficiently. This strongly suggests that we identify quasar broad emission lines with clouds driven away from the disk \citep{emmering92,konigl94,bottorff97}, an interpretation that should be testable. By contrast, in the BLL, the low density of optical and infrared photons allows protons to be accelerated to higher energy, emitting TeV photons which can now escape from small radii without strong shielding. The discovery of TeV emission from FSRQ with minute-scale variability would exclude this model.

\subsection{Environmental Impact}\label{ssec:envimp}
AGN jets heat their galactic surroundings, efficiently in the case of FR-I sources and rather inefficiently in the case of the more powerful FR-II sources, which, instead, create hot cocoons that help protect jets from destructive instability \citep{fabian12}. After the jet switches off or declines in power, these cocoons may detach to form giant bubbles that rise buoyantly away from the galaxy. They are ultimately assimilated by the circumgalactic medium, which they heat. This is most important when the host galaxies reside in a rich galaxy cluster. Jets can have a more immediate effect by stimulating star formation as is sometimes seen in some galaxies and, more impressively, by triggering the formation of new galaxies as the jets can propagate several Mpc away from their hosts. Jets also accelerate high energy cosmic rays and quite plausibly may account for a large fraction of the universal spectrum above the ``knee'' in the spectrum. They may also be responsible for most of the intergalactic magnetic field. 

So, AGN play a major role in the evolution of the universe, arguably comparable to that played by stars, and the jets can mediate this interaction directly. Quantitative measures of this role are still quite uncertain and must be compared with the influence of the radiation and outflowing winds that are associated with the accretion disks. All of this must be considered in the young universe when most nuclear activity occurred and massive black holes grew, and even earlier when the intergalactic medium was reionized and the first stars and galaxies were formed. We, therefore conclude this article with a sketchy overview of attempts to address some of these issues and to relate clear taxonomic structure in the observed properties of radio sources to underlying physical processes.  

\subsubsection{Cosmological Evolution of Extragalactic Jets}\label{ssec:cosev}
Ever since radio sources were first identified with optical sources (galaxies and quasars), of moderate and high redshift, it has been obvious that the radio source population was quite different at early cosmological times than at present \citep{pooley67,schmidt68,rees71}.  As beautifully shown by \citet{wall77}, when source counts, local radio luminosity function, and redshift data are combined, it is clear that it is the bright (FR-II) end of the luminosity function that was much more populous in the past.  At the present epoch powerful radio sources and quasars are rather rare.  

As shown above, the jet powers generated by both accretion disk and rotating black hole are an increasing function of the black hole feeding rate $\dot{m}$. Furthermore, a large body of work in the fields of coevolution of black hole and galaxy growth have tied the star formation rate closely to the long term feeding rate of the central hole, both of which are believed to be driven ultimately by the rate of mergers of galaxies with one another \citep{ho04,padovani16}.  A plausible explanation for the decrease in activity of powerful radio sources over time, therefore, is the decrease in galaxy merger rate over that same time. Much work has been done on computational studies of large cosmological volumes of galaxy formation and merging.  However, given all the complex processes at work, our understanding of the evolution of AGN jets will continue to be led by the results of large surveys. 

\subsubsection{Radio Source Lifetimes and Duty Cycle}\label{ssec:life}
The term ``lifetime'' can refer to at least three different time scales for radio sources: 1)  The radiative lifetime is the period in which all the energy in the source lobes would be radiated away at its present luminosity.  2)  The engine lifetime is the period during which the black hole spins rapidly and/or it is being fueled sufficiently. The absence of ``orphaned'' lobes  (\textbf{Section~\ref{sssec:rapvar}}) implies that the engine lifetime must be somewhat longer than the lobe radiative lifetime in most sources. 3)  The cycle time is the period between engine outbursts for given radio source, during which jet production is dormant.  

It is probably the case that each radio galaxy goes through different luminosity stages in its history.  For example, M87 was probably a powerful FR-II galaxy in early cosmic times, but eventually became the weak FR-I radio galaxy it is today.  Therefore, actual values of the three above time scales make sense only when specified for a given radio source class.  For example, the radiative lifetime of FR-II sources is $\sim 10^{6-7}$ yr (O'Dea 08), while the maximum engine lifetime (as measured by estimating the maximum FR-II lobe expansion and dissipation time in galaxy groups and clusters) is $\sim 1.7 \times 10^7$ yr \citep{bird08}.  These authors also estimated the time between each engine-fueling episode to be roughly $\sim 10^9$ yr.   So, powerful radio source engines seem to produce a jet for $\lesssim 17$ Myr, but can be re-triggered every 1 Gyr or so, with a duty cycle of $< 2$\%.  

In the case of most FR-I sources, the number density of these low power radio galaxies does not appear to evolve much with cosmic time \citep{schmidt72}.  That is, the population evolutionary time scale for FR-Is is very long -- greater than the age of the universe. This suggests that the engine fueling may be essentially continuous for FR-Is, but at a low rate.  

\section{Summary}\label{sec:sum}
\begin{extract}
``It's an exercise in delayed gratification.'' -- \citet{doeleman17}
\end{extract}
\subsection{Towards a Working Model of AGN Jets}\label{ssec:wmod}
Relativistic jets are the conduits that connect supermassive, spinning black holes and their attendant disks to their host galaxies and the universe beyond. They were once seen as exhausts that, like their automotive counterparts, remove excess heat from powerful machines. However, the observations that we have reviewed suggest a rather different metaphor. Black holes are turbines that are spun up by orbiting gas to generate high voltage electrical power and AGN jets are lossy and glowing, coaxial cables that ultimately heat their surroundings. This change of viewpoint is one that is supported by observations of selected local AGN, from which we tentatively generalize to AGN in general. However it is not secure. 

Nonetheless, this sketch  does suggest causal and testable mechanisms through which the type of AGN that is produced is determined by the black hole mass, spin, gas supply rate, and the strength of the magnetic field, which, in turn may depend upon the manner by which gas is supplied to the nucleus at $R_{\rm inf}$. In particular, it appears that only powerful jets, can escape the nucleus as supersonic, relativistic flows. It also indicates that AGN jets can accelerate protons to EeV energy and that they could be significant sources of high energy cosmic rays. This further allows jet $\gamma$-ray emission to be mostly synchrotron instead of Compton radiation. \textit{In toto}, AGN contribute to and are constrained by the $\gamma$-ray background \citep{dimauro18}. Understanding the content, power, electrical current and duty cycles of AGN jets will help quantify their role in galaxy formation and evolution.  

\subsection{Observing AGN Jets with New Telescopes}\label{ssec:ntel}
These are exciting times in the study of AGN jets. Our understanding of how they operate and how they relate to other manifestations of nuclear activity has matured over the past decade through the opening up of the $\gamma$-ray spectrum, the development of mm VLBI and the systematic, multi-wavelength study of large samples. High dynamic range RMHD and PIC simulations have also solidified our basic physical understanding of how black holes and their disks release energy and how their coronae, winds and jets accelerate particles to impressive energies. While many questions remain, the prospects are good for giving them clean answers soon because several new observational facilities are coming on line.

The most immediate of these is EHT. This is making mm, and eventually attempting submm, images with limiting angular resolution of a few gravitational radii in the best-resolved cases led by M87 and SgrA$^\ast$ so that general relativistic optics will be necessary to measure the black hole masses and spins and to interpret the images. For M87, it should settle the question of whether jets are powered by the hole or the disk and, through its polarimetric capability, demonstrate if the initial jet collimation is due to gas pressure exerted by a thick disk or wind or magnetic stress. The thickness of the SgrA$^\ast$ disk should also be revealed and a weak jet will be either detected or bounded. A larger sample of sources will be used to determine if the model of the M87 black hole jet region is more widely applicable. Similar scales are being probed by the GRAVITY instrument deployed on the VLT which has reported strong astrometric and polarimetric evidence for features orbiting the SgrA$^\ast$ black hole close to the ISCO \citep{abuter18}.

In the optical infrared band, James Webb Space Telescope (2021), Giant Magellan Telescope (2023), Extremely Large Telescope (2024), Thirty Meter Telescope (2027) will observe the inner parts of galaxies with angular resolution of $\sim100\,{\rm mas}$ at $2\,\mu$. JWST will have a far greater sensitivity but ground based telescopes can improve the angular resolution by nearly an order of magnitude using adaptive optics. Together these telescopes should be able to resolve Galaxy Jets and define both the gravitational potential and the interstellar medium within $R_{\rm inf}$ through which they propagate. They will also help understand the flow of gas, both inward and outward, and how it relates to accretion as well as reveals some of the impact a powerful jet can have on its environment.

The Cerenkov Telescope Array (CTA) comprises two observatories in Chile and La Palma and should begin construction in 2019 to be fully operational in the middle of the next decade. It will have ten times the sensitivity of existing telescopes, an improved energy coverage from $\sim30\,{\rm GeV}$ to $\sim100\,{\rm TeV}$, angular resolution up to $\sim3'$ at high energy and much better time coverage of the most prominent sources. It should greatly improve our understanding of jet beaming and variability and help us determine the geometry, speeds and emission mechanisms of black hole jets. LHAASO will use water Cerenkov detectors to extend the gamma range to $\sim1\,{\rm PeV}$. 

In the X-ray band, the Imaging X-ray Polarimetry Explorer (IXPE) should be able to detect many more X-ray blazars and learn about their magnetic field structure and, on a longer timescale, ATHENA will improve on existing telescopes in sensitivity and resolution.

The long term future of radio astronomy is defined by the Square Kilometer Array (SKA) project and the next generation Very Large Array (ngVLA) proposal. Collectively these undertakings, with their precursors and pathfinders, advertise up to fifty times the sensitivity of existing radio telescopes from 50 MHz to 116 GHz. They will also greatly improve the sensitivity of cm wavelength VLBI. They will be able to observe very large samples of extragalactic radio sources and map their black hole jets so as to develop a much better statistical understanding of beaming and structure on pc scales.

Many more telescopes will be combined with these facilities, working in multi-wavelength/messenger mode. Collectively, they are confidently expected to make discoveries that will advance the scientific program from the following list of goals tentatively suggested right now by this review.

\begin{issues}[FUTURE ISSUES]
\begin{enumerate}
\item Discover about gas flow and jet production around black holes using mm and submm EHT observations
\item Understand if the sustenance of magnetic field near a black hole, which determines whether an AGN is radio-loud or radio-quiet is due to physical processes near the black hole or is controlled by the infalling gas
\item Employ VHE neutrino observations to open up a new window on AGN and determine if blazar jets are major cosmic ray sources
\item Use EHT to study many jets on the scale of $R_{\rm inf}$ and verify  that the FR class of a radio source is determined here. 
\item Learn how to map emission in a given spectral band onto jet radius.
\item Determine if the TeV emission from blazars is synchrotron or Compton radiation
\item Develop hybrid numerical simulations that meld relativistic hydromagnetic descriptions with particle kinetics and radiative transfer.
\item Quantify the role of AGN jets in promoting and limiting galaxy formation and evolution. 
\end{enumerate}
\end{issues}

\section*{DISCLOSURE STATEMENT}
The authors are not aware of any affiliations, memberships, funding, or financial holdings that might be perceived as affecting the objectivity of this review. 
\section*{ACKNOWLEDGMENTS}
We thank Ramesh Narayan for his careful and constructive review of the first draft of this article and Noemie Globus for a careful reading of the revised draft. Beyond this, our debt and gratitude to more colleagues than we can possibly list, who have framed our evolving views on AGN jets over the past fifty years, are deep and sincere.

\bibliographystyle{ar-style2}

\end{document}